\documentclass[fleqn,usenatbib]{mnras}

\usepackage[T1]{fontenc}

\DeclareRobustCommand{\VAN}[3]{#2}
\let\VANthebibliography\thebibliography
\def\thebibliography{\DeclareRobustCommand{\VAN}[3]{##3}\VANthebibliography}

\usepackage{graphicx}	
\usepackage{amsmath}	
\usepackage{amssymb}	

\usepackage{newtxtext,newtxmath}

\usepackage{subfig}
\usepackage{multirow}
\usepackage{tabularx}
\usepackage[normalem]{ulem}
\usepackage{array}
\usepackage[usenames,dvipsnames]{xcolor}
\usepackage{breqn}
\usepackage{booktabs} 
\usepackage{makecell}
\usepackage{lastpage}
\usepackage[final]{microtype}

\newcommand{\be}{\begin{equation}}
\newcommand{\ee}{\end{equation}}
\newcommand{\mach}{\mathcal{M}}

\title[Evolution of AGN stars]{In-situ extreme mass ratio inspirals via sub-parsec formation and  migration of stars in thin, gravitationally unstable AGN discs } 
\author[A. Derdzinski \& L. Mayer]{Andrea  Derdzinski$^{1}\thanks{E-mail: andrea@ics.uzh.ch}$ and Lucio Mayer$^{1}$  \\
$^{1}$Center for Theoretical Astrophysics and Cosmology, Institute for Computational Science, University of Zurich,
Winterthurerstrasse 190, 8057 Zurich, Switzerland\\}

\begin{document}
\date{Received / Accepted}
\pagerange{\pageref{firstpage}--\pageref{lastpage}} \pubyear{2020}

\maketitle
\label{firstpage}

\begin{abstract}
We investigate the properties of stars born via gravitational instability in  accretion discs around supermassive black holes (SMBHs) in active galactic nuclei (AGN), and how this varies with the SMBH mass, accretion rate, or viscosity. We show with geometrically thin, steady-state disc solutions that fragmentation results in different populations of stars when one considers the initial conditions (e.g. density and temperature of the gravitationally unstable regions).  We find that opacity gaps in discs around $10^6 \rm M_{\sun}$ SMBHs can trigger fragmentation at radii $\lesssim 10^{-2}$ pc, although the conditions lead to the formation of initially low stellar masses primarily at $0.1-0.5 \rm M_{\sun}$. Discs around more massive SMBHs ($M_{\rm BH} =10^{7-8} \rm M_{\sun}$) form moderately massive or supermassive stars (the majority at $10^{0-2} \rm M_{\sun}$). Using linear migration estimates, we discuss three outcomes: stars migrate till they are tidally destroyed, accreted as extreme mass ratio inspirals (EMRIs), or leftover after disc dispersal.  For a single AGN activity cycle, we find a lower-limit for the EMRI rate  $R_{\rm emri}\sim 0-10^{-4} \rm yr^{-1}$ per AGN assuming a SF efficiency $\epsilon=1 -30\%$. In cases where EMRIs occur, this implies a volumetric rate up to $0.5-10 \rm yr^{-1} Gpc^{-3}$ in the local Universe. The rates are particularly sensitive to model parameters for $M_{\rm BH}=10^6 \rm M_{\sun}$, for which EMRIs only occur if stars can accrete to $10$s of solar masses. Our results provide further evidence that gas-embedded EMRIs can contribute a substantial fraction of events detectable by milliHz gravitational wave detectors such as LISA. Our disc solutions suggest the presence of migration traps, as has been found for more massive SMBH discs.  Finally, the surviving population of stars after the disc lifetime leaves implications for stellar discs  in galactic nuclei.

\end{abstract}


\section{Introduction} 
\label{sec:introduction}

The most popular interpretation of quasars proposes that supermassive black holes (SMBHs) with masses $M_{\rm BH}\gtrsim 10^{5} {\rm M_{\sun}}$ 
reside in galactic nuclei surrounded by dense gaseous accretion discs. 
The bright emission originates from the accretion of gas onto the SMBH, which is driven by some mechanism of viscous dissipation. 
While the precise structure of the accretion flow onto 
SMBHs is poorly constrained and may take many forms,  
models of geometrically thin, near-Keplerian discs (e.g. \citealt{SS1973})
reproduce observed continuum emission in several cases 
leading such geometry to be considered  a standard benchmark for accretion flow in quasars \citep{Krolik1999}. 
Similar models explain active galactic nuclei (AGN) throughout the Universe to varying degrees, particularly those that are bright or shining at near-Eddington rates. 
An inevitable implication of thin disc models, however,  is their 
vulnerability to gravitational instability in the outer regions (e.g. \citealt{Shlosman1987,Goodman2003}).

Gravitational instability (GI), initially introduced in the context of stellar discs by \cite{Toomre1964}, 
occurs when the self-gravity of the gas overcomes its pressure and rotational support. 
Depending on the ability of the gas to cool, the gravitationally unstable disc can fragment into coherent structures, leading to the formation of bound clumps \citep{Gammie2001},  possibly halting the accretion flow entirely \citep{1989ApJ...341..685S,Shlosman1990,InayashiHaiman2016}. 
GI also generates spiral density waves \citep{2007prpl.conf..607D}. 
These nonaxisymmetries can dominate the angular momentum transport in heavily self-gravitating discs (e.g. \citealt{Kumar1999,Cossins2009,LodatoRice2004,LodatoRice2005}) and ultimately determine 
 the disc evolution \citep{Boley2010}. 
In turn, energy dissipation due to gas collapse and density wave propagation generates heat, which can re-stabilise the disc. 
The interplay between collapse and restabilization suggests that regions of discs may exist in a quasi-steady state of inflow, or a \emph{gravitoturbulent} state in the  case that fragmentation is suppressed \citep{Rafikov2009}, where the accretion flow is partially driven albeit not terminated by GI. 

Fragmentation due to marginal gravitational instability has been proposed as a mechanism for 
star formation in AGN discs \citep{Shlosman1987,WangSilk1994,Levin2007,Goodman2003,GoodmanTan2004,Nayakshin2006} that may explain stellar discs in the Galactic center \citep{LevinBeloborodov2003,Nayakshin2005,Davies2020}, in addition to altering the growth rate of SMBHs \citep{DittmannMiller2020}. 
Analogously it is considered to be a formation mechanism for planets in gravitationally unstable protoplanetary discs \citep{Boss1997,Helled2014}. 
Star formation in AGN discs and subsequent growth and interactions of embedded compact remnants (CRs) can lead to interesting observational signatures in both electromagnetic emission and gravitational waves (GWs) across a range of frequencies. 
AGN-assisted CR interaction is a proposed formation channel for high-frequency GW events detectable by LIGO-Virgo, in particular to explain detected stellar origin BH-BH mergers with large constituent masses and events with unequal mass ratios or unusual spins \citep{Bartos2017,Stone2017, Secunda2019,Yang2019,McKernan2018,McKernan2020,Tagawa2020a,Tagawa2020b,Tagawa2021b,Gayathri2021}.
Embedded CRs (or stars, if not disrupted) may also migrate to the central regions of the disc where they interact with the central SMBH and become lower-frequency GW sources, referred to as extreme- or intermediate- mass ratio inspirals (E/IMRIs), which will be  
detectable by future space-based interferometers such as LISA \citep{Amaro-Seoane2017}, TianQin \citep{Fan2020}, or TAIJI \cite{Gong2021}.  

EMRIs are a prime target for milliHz GW missions. The detection of even one will provide high precision measurements of the central SMBH mass and spin \citep{2007PhRvD..75d2003B,2017PhRvD..95j3012B}. If the EMRI is embedded in gas, the GW signal can carry signatures of the environment that can be potentially constrain environment properties and complex migration physics  \citep{Yunes2011,Kocsis2011,Barausse2014,Derdzinski2019,Derdzinski2021,Zwick2021}.
Rate estimates of EMRI formation via dynamical interactions in galactic nuclei span orders of magnitude in uncertainty \citep{2020arXiv201103059A}. Luckily, it is becoming apparent that the presence of an accretion disc can boost the EMRI formation rate \citep{PanYang2021a,PanYang2021b}. One mechanism for EMRI facilitation occurs due to the influence of the accretion disc on the orbits of stars and CRs in the nucleus, which tend to align and corotate with the disc \citep{1991MNRAS.250..505S,1983ApJ...266..502N,Fabj2020, Tagawa2020a}. 
Additionally, instability-induced star formation can produce in-situ CRs that  accrete, merge, or migrate towards the central SMBH \citep{Levin2007,GoodmanTan2004}. The latter works describe the collapse and aggregation of clumps in marginally stable discs that leads to the formation of a single, supermassive embedded star.

In this work, we revisit models of gravitationally unstable accretion discs, extending solutions over a set of system parameters (namely SMBH mass and rate of viscous inflow), to demonstrate the range of initial conditions for in-situ AGN stars. 
We investigate the diversity of the stellar mass distribution and subsequent evolutionary outcomes. Not all AGN are the same\textemdash SMBH masses, the total gas mass, and efficiency of inflow can vary, leading to different predictions for the location of fragmentation and the AGN star evolutionary outcomes: tidal disruption, in-situ EMRIs, or leftover stellar populations. 
In the following Section, we review recent developments of fragmentation in the protoplanetary disc context, for which several numerical investigations 
find deviations from the standard theory when attempting to explain properties of planets formed through GI.
We then briefly summarize theoretical aspects of SF employed in this work, before describing our disc model in Section~\ref{sec:discmodel}. 
 With solutions for disc structure, we derive the regime where fragmentation is expected to occur in Section~\ref{sec:fragmentation}.
In Section~\ref{sec:massesofstars} we describe the process of fragmentation to gravitational collapse of protostars, based on the size of the instability, the collapse rate, and the accretion onto the resulting cores, and we derive the initial mass distribution of stars. 
In Section~\ref{sec:evolution} we describe the subsequent evolution of AGN-star populations, given their ability 
to migrate and evolve throughout the disc.  
In Section~\ref{sec:outcomes} we discuss the outcomes of embedded stars, which can be tidally disrupted or accreted by the SMBH, and we quantify the rate of in-situ EMRIs for each disc solution. 
Finally, in Section~\ref{sec:discussion} we summarise the takeaways and caveats of this work.

\section{Initial conditions}

\subsection{Astrophysical rationale}
\label{sec:planetsstars}

The criteria for fragmentation of gas into bound objects has been extensively tested in numerical simulations, often in the context of protoplanetary discs. 
It is well understood that fragmentation will occur if the disc cools on a timescale comparable to the local orbital time \citep{Gammie2001,Mayer2008,JohnsonGammie2003,Clarke2007}, 
although determining the precise conditions has proved a challenging numerical problem. 
The criteria for fragmentation, as well as the outcome of collapsing clumps (whether at the scale of planets,  stars, or something in between) are sensitive to numerical implementations of 
 disc thermodynamics \citep{Rice2005,Mayer2004},
 numerical viscosity \citep{Hu2014},
gravitational softening \citep{YoungClarke2015},
and small-scale resolution \citep{Nelson2006,Paardekooper2012,Brucy2021}.
Recent numerical studies find convergence (over implementations of artificial viscosity) on the critical cooling rate below which fragmentation occurs, finding $\beta_{\rm crit} \equiv t_{\rm cool} \Omega \approx 3$ \citep{Deng2017}, where $t_{\rm cool}$ is the radiative cooling timescale and $\Omega = (G M/r^3)^{1/2}$ is the orbital frequency at a radius $r$ in a disc around a central mass $M$. This is the criteria we adopt in this work.

Convergence on the size and mass distribution of resulting clumps has yet to be attained numerically, however.
Linear perturbation theory that computes the growth rate of local axisymmetric density perturbations in razor-thin discs
predicts that fragmentation produces clumps on a characteristic scale, namely the most unstable Toomre wavelength \citep{Toomre1964}.
Yet several simulations find that clumps form at systematically  smaller sizes, even differing by an order of magnitude from the predictions
of perturbation theory
(e.g. \citealt{Boley2010,Galvagni2012,Galvagni2014,Muller2018,Deng2021,2011MNRAS.417.1928F}).
Differences arise due to finite disc thickness, departure from axisymmetry, and nonlinear
effects in the gas flow dynamics immediately prior to fragmentation.
In particular, fragmentation is almost invariably seen to occur along 
spiral arms generated in the initial phase of a gravitational instability in realistic discs on a variety of scales, 
hence clumps form in a gas flow having inherently a non-axisymmetric character (see, e.g. \citealt{Boley2010} for protoplanetary disc scales or \citealt{2015MNRAS.453.2490T} for galaxy-scale fragmentation). In addition, soon after fragmentation ensues, other effects matter in
the nonlinear stage, which, by construction, cannot be captured by perturbation theory, such as the rotational support in clumps \citep{Galvagni2012}, and  their vulnerability by local shear and
stellar tidal forces \citep{Galvagni2014,Muller2018}. 
Furthermore, and perhaps most critical for ionised gas in AGN discs, large deviations from predictions of  Toomre instability theory, both before
and after fragmentation, arise in simulations of discs threaded with magnetic fields. This results in  clump sizes that are even two orders of 
magnitude smaller than those in non-magnetized discs 
\citep{Deng2020}.
When estimating the characteristic sizes of pre-stellar clumps in this work, we draw from the results of recent numerical simulations to take these reductions into account. \\


Considering that we are concerned with fragmentation at the AGN disc scale, 
we must also consider basic details of star formation. 
Conventional star formation involves the collapse of diffuse gas to dense, accreting cores, involving a vast range of temperatures and physical scales. 
Determining the stellar mass distribution from a collapsing cloud is a non-trivial problem given that complexities in opacity, radiative processes, rotation, and magnetic fields are all critical for determining the resulting stellar masses \citep{Larson1969,Penston1969,StahlerPalla2004}, particularly in regimes of high accretion rates that lead to supermassive stars \citep{Haemmerle2021, Hokosawa2019}.

In the case of star formation induced by GI in an AGN disc, fragmentation occurs in high angular momentum material and at 
initially higher densities and temperatures than a typical molecular cloud.
Protostars 
effectively skip the initially slow phase of prestellar contraction\footnote{
The initial collapse of an optically thin cloud at $T_0\sim10$s K, known as the `first core' phase, proceeds until the central core reaches H dissociation temperatures ($\sim 2000$ K). Beyond this stage the protostar continues to grow as a stable object in the `main accretion' phase \citep{StahlerPalla2004}. AGN pre-stellar cores are born in the main accretion phase.}.
Such cores collapse rapidly due to the gas conditions, as we will show, experiencing
high accretion rates. Certain conditions lead to the formation of supermassive stars (SMSs), for which there are greater uncertainties in the final stellar masses and structure \citep{Hosokawa2013} and subsequent evolution \citep{Haemmerle2021}. 
Critically, the formation and subsequent evolution of AGN-stars does not occur in isolation:  gas from the disc may feed the protostellar envelopes and exert torques.

\subsection{Disc model}
\label{sec:discmodel}
We define a system of equations describing a thin, Keplerian, steady-state accretion disc. 
The central BH accretes at a constant fraction of the Eddington rate $\dot{M} = f_{\rm Edd} \dot{M}_{\rm Edd}$, where the Eddington rate $\dot{M}_{\rm Edd} = \frac{4 \pi G M_{\rm BH}}{\epsilon_{\rm eff} \kappa_{\rm es} c}$ for a central BH of mass $M_{\rm BH}$ and a radiative efficiency $\epsilon_{\rm eff} = 0.1$. 
The viscosity is parameterised with the alpha-disc prescription, as in \citet{SS1973}. 
The disc model is similar to that presented in \citet{SirkoGoodman2003}, except that, while they adopted tabulated opacities (from \citealt{IglesiasRogers1996} and \citealt{AlexanderFerguson1994}), we use analytical opacity fits by \citet{BellLin1994}.

\begin{figure*}
\begin{center}
\includegraphics[width=0.97\textwidth]{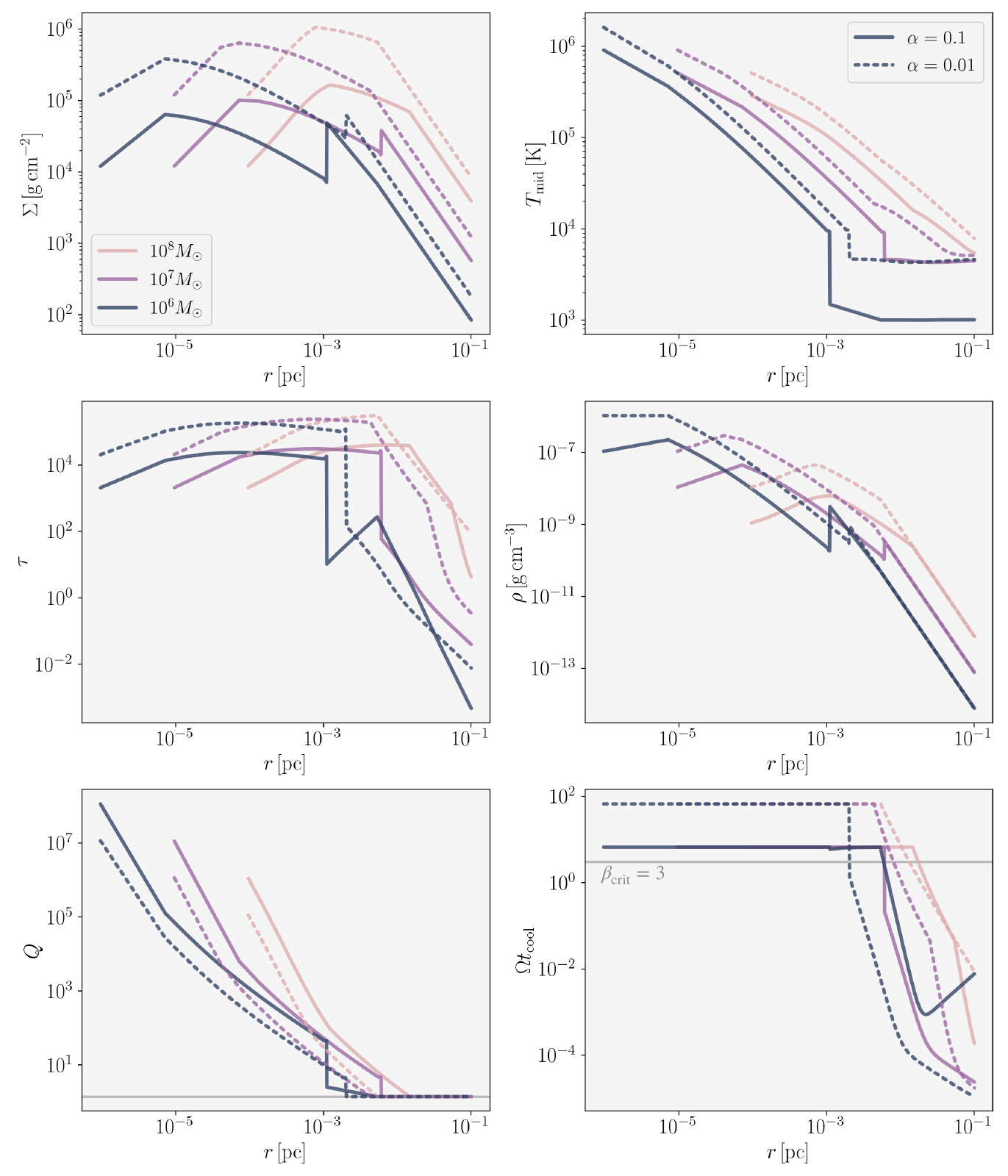}
\caption{Solutions for the alpha disc model around SMBH masses $10^6 {\rm M_{\sun}}$, $10^7 {\rm M_{\sun}}$, and $10^8 {\rm M_{\sun}}$ (dark to light lines, respectively), accreting at $\dot{M}/\dot{M}_{\rm Edd} = 0.1$ with $\alpha=0.1$ (solid lines) and $\alpha=0.01$ (dashed lines). [\emph{Top panels}]: surface density (left) and midplane temperature (right). [\emph{Middle panels}]: optical depth (left) and density (right). [\emph{Bottom panels}]: Toomre Q (left) and the ratio of cooling time to dynamical time (right). Discontinuous changes in profiles are an artifact of a piece-wise opacity law. Discs around less massive SMBHs show a sharp density transition at $r\sim10^{-3}$ pc. This is due to a drop in opacity as a result of H- recombination. A lower viscosity disc allows for higher densities, which alters the location of opacity transitions.}
\label{fig:num_discs}
\end{center}
\end{figure*}

\begin{table*}
\begin{tabular}{lcccr@{\hskip 1mm}c@{\hskip 0.05 mm}l}\\
Opacity regime  & $\kappa_0 \, \rm [cm^2 g^{-1}]$  & $a$ & $b$ & \multicolumn{3}{c}{Temperature range [$K$]} \\
\specialrule{0.8pt}{1pt}{1pt}
\vspace{-0.3cm}
\\
ices  & $2\times 10^{-4}$  & $0$ & $2$  & & $T$ & $\le 166.810$ \\
ice sublimation  & $2\times 10^{16}$  & $0$ & $-7$  & $166.810<$ &$T$ & $\le 202.677$ \\
dust grains  & $1\times 10^{-1}$  & $0$ & $1/2$  & $202.677<$ & $T$ & $\le 2286.77 \rho^{2/49}$ \\
dust sublimation  & $2\times 10^{81}$  & $1$ & $-24$  & $2286.77 \rho^{2/49} <$& $T$& $\le 2029.76 \rho^{1/81}$ \\
molecules  & $1\times 10^{-8}$  & $2/3$ & $3$  & $2029.76 \rho^{1/81}<$& $T$& $\le 10^4 \rho^{1/21}$ \\
H- recombination  & $1\times 10^{-36}$  & $1/3$ & $10$  & $10^4 \rho^{1/21}<$ &$T$& $\le 31195.2 \rho^{4/75}$ \\
bound-free + free-free  & $1.5\times 10^{20}$ & $1$ & $-5/2$ & $31195.2 \rho^{4/75}<$ & $T$& $\le 1.79393\times10^8 \rho^{2/5}$ \\
e$^-$ scattering  & $0.348$  & $0$ & $0$  & & $T$&$>1.79393\times10^8 \rho^{2/5}$ \\
\specialrule{0.8pt}{1pt}{1pt}
\end{tabular}
\caption{Piecewise opacity scalings of the form $\kappa = \kappa_0 \rho^a T^b$ and corresponding temperature ranges from \citet{BellLin1994}. Given that in regions of low densities multiple conditions can be satisfied, we also set the constraint that $T>2\times10^{-4} \rm \, K$ for electron scattering.}
\label{table:belllin}
\end{table*}

The effective temperature is determined from internal heating via viscous dissipation, and scales with $\dot{M}$:
\be
\label{eq:teff}
T_{\rm eff}^4 = \frac{3}{8 \pi \sigma} \Omega^2 \dot{M'}
\ee
where $\sigma$ is the Stefan-Boltzmann constant, $\Omega = (G M / r^3)^{1/2}$ is the orbital frequency and $\dot{M'} = \dot{M}(1-(r_{\rm min}/r)^{1/2})$ arises from applying a zero-torque boundary condition at the disc inner radius $r_{\rm min}$.
Assuming the disc is in steady state with a constant mass flux, the surface density is given by 
\be
\Sigma = \frac{\dot{M'}}{3 \pi \nu}
\label{eq:sigmanu}
\ee
where $\nu = \alpha c_s^2 \Omega^{-1}$ is the kinematic viscosity, assuming that viscosity is limited by disc-scale turbulence,\footnote{Turbulence may be driven by magneto-rotational instability, which is expected to occur in a rotating ionized gas and is considered one of the main mechanisms for angular momentum transport in inner AGN discs.}
and $\alpha<1$.  
The disc is described by a scale height that depends on the sound speed:
\be
h = \frac{c_s}{\Omega}
\ee
which also determines the structure and midplane density at each radius by
\be
\rho= \frac{\Sigma}{2 h}.
\ee
The mid-plane temperature is derived from the diffusion equation, assuming that photons generated in the midplane are transported radiatively to the disc surface according to an effective opacity $\tau_{\rm eff}$:
\be
T_{\rm mid}^4 = \tau_{\rm eff} T_{\rm eff}^4.
\label{eq:diffusion}
\ee
where  $\tau_{\rm eff} = \frac{3}{8}\tau + \frac{1}{2} + \frac{1}{4 \tau} $ interpolates between the optically thick and thin regimes, 
and we have used the Eddington approximation for the photospheric boundary, assuming that the gas is in thermal equilibrium. 
 $\tau$ is the optical depth:
\be
\tau = \frac{1}{2} \kappa \Sigma,
\ee
computed with the Rosseland mean opacity of the gas at the disc midplane, that is given by a piecewise function following the form
\be
\label{eq:opacity}
\kappa = \kappa_0 \rho^a T_{\rm mid}^b
\ee
where $\kappa_0$, $a$, and $b$ are defined over regimes of $\rho$ and $T_{\rm mid}$ following the fits from \citet{BellLin1994}. The opacity laws include regimes from ice, dust grains, molecular absorption, Hydrogen recombination, to free-free and bound-free absorption and electron scattering. Opacity scalings and piecewise temperature ranges are summarized in  Table~\ref{table:belllin}.

The sound speed depends on both gas and radiation pressure 
\be
c_s^2 = \frac{p_{\rm gas}+p_{\rm rad}}{\rho} = \frac{k_{\rm B}}{\mu m_{\rm H}} T_{\rm mid} 
+ \frac{\sigma_{\rm B}}{2 c} \frac{\tau T_{\rm eff}^4}{\rho},
\ee
where $k_{\rm B}$ is the Boltzmann constant, $m_{H}$ is the mass of Hydrogen, $c$ is the speed of light, the mean molecular weight is taken to be $\mu = 0.62$ for an ionized gas. For $p_{\rm rad}$ we use an effective radiation pressure as in \citet{SirkoGoodman2003}. In the limit of high opacity ($\tau\gg1$), this reduces to the typical scaling with midplane temperature ($p_{\rm rad}=4 \sigma T_{\rm mid}^4/(3 c) $). When the gas is optically thin ($\tau<1$), $p_{\rm rad}\propto \tau^2 T_{\rm mid}^4$, and the extra opacity dependence arises as a consequence of the inefficiency of radiation in an optically thin medium.

This gives us effectively a system of 7 equations with 7 variables:  $\rho(r)$, $\Sigma(r)$, $T_{\rm mid}(r)$, $\kappa(r)$, $\tau(r)$, $h(r)$, and $c_s(r)$, which can be solved numerically at each radius $r$ given a choice of central mass $M$, accretion rate $\dot{M}$ (which implies  $T_{\rm eff}(r)$), and viscosity parameter $\alpha$.

Once solving the above equations over a range of radii, we can calculate the Toomre stability parameter $Q$ \citep{Toomre1964, Goldreich1965}:
\be
Q = \frac{c_s \Omega}{\pi G \Sigma}.
\label{eq:toomreq}
\ee
Beyond the radius where the Toomre $Q$ becomes less than a critical value $Q_0\equiv1.4$, the disc is gravitationally unstable.  
In this regime, we restrict $Q=Q_0$, effectively holding the disc in a marginally stable, self-regulated state. The solutions for the outer disc carry simple, analytical scalings with chosen parameters, as shown in several previous works (\citealt{Rafikov2009,Rafikov2015,Nayakshin2006}, among others)
The assumption of a constant $\dot{M}$ implies a constant midplane temperature and sound speed, and in turn changes the condition for the outer disc density to be 
\be
\label{eq:rho}
\rho = \frac{\Omega^2}{2 \pi G Q_0}.
\ee
Mass conservation means that Eq.~\ref{eq:sigmanu} still applies, although the source of viscosity mediated by $\alpha$ should be no longer interpreted as the scale of turbulent eddies but by the effectiveness of nonaxisymmetries due to GI. 
 
The cooling time of the disc is calculated by 
\be
t_{\rm cool} = \frac{1}{\gamma - 1} \frac{3}{16} \frac{\Sigma c_s^2}{\sigma_b T^4} \left( \tau + \frac{1}{\tau} \right),
\ee
where we simply take $\gamma = 5/3$ as the adiabatic index. Here we again utilize the opacity at the midplane of the disc, given that our model does not explicitly solve for the vertical disc structure.
This estimate for the cooling time comes from the presumption that the energy dissipation due to viscous transport must equal the energy flux through the surface of the disc. Thus, from the equations above, $t_{\rm cool}\propto\Omega^{-1}$ in the inner disc (ensuring steady-state) and drops (in a manner dependent on the opacity) in the outer disc once the assumption of viscous dissipation (via Eq.~\ref{eq:teff}) breaks down with the onset GI. 

By assuming the viscosity is dependent on the total pressure, the radiation-pressure dominated region of the inner disc is thermally unstable \citep{1978ApJ...221..652P,Pringle1981}. One workaround for this is to assume the viscosity depends on the gas pressure only (the so called beta-disc model, \citealt{LyndenBellPringle1974,Hure2001}), which will weaken the viscosity in the inner region and (to maintain the same $\dot{M}$) lead to a higher density. 
For this work we stick with the alpha-disc assumption, given that it does not strongly affect the properties in the outer disc region where gas pressure is dominant.

Full disc solutions are shown for a range of SMBH masses and two values of $\alpha$ in Fig.~\ref{fig:num_discs}. One prominent feature in our disc solutions for lower SMBH masses is an opacity gap at $\sim10^{-3} \rm pc$. 
As we decrease the central SMBH mass, the lower densities and temperatures in the disc allow for the dominant opacity to change from electron scattering and bound-free/free-free transitions to H- recombination, resulting in a sharp opacity drop and a corresponding change in the density profile (further discussed in Section~\ref{sec:fragmentation}). 
Note that the adoption of piecewise opacity powerlaws\footnote{While the \citet{SirkoGoodman2003} disc solutions show multiple solutions for disc temperature (due to the use of opacity tables), our simplified opacity laws lead to only a single solution for the temperature profile.} results in unrealistically sharp transitions in the disc profiles. In a more realistic system, these transitions should be smoother given that opacity regimes may overlap. 

We also consider the effect of a temperature background on the disc, which can increase $T_{\rm eff}$ and affect the cooling rate. There is observational evidence that surface irradiation contributions a substantial fraction of the disc temperature from reverberation mapping studies of some Seyfert AGN \citep{Fausnaugh2016,Starkey2017}. If we assume that the temperature background is sourced by the accretion luminosity emitted from the central BH (or the inner disc region), then Eq.~\ref{eq:teff} obtains an additional component due to external disc heating: 
\begin{equation}
T_{\rm eff, irr}^4 =  \frac{3}{8 \pi \sigma} \Omega^2 \dot{M'} 
+ \frac{\epsilon_{\rm eff} \dot{M} c^2 }{4 \pi \sigma r^2} \frac{h}{r}.
\end{equation}
Here we assume the disc albedo is zero, meaning that all incident radiation is absorbed, in order to fully demonstrate the effect of irradiation. We find that incorporating this term increases $T_{\rm eff}$ by a factor of a few, and has the strongest effect for the $M_{\rm BH} = 10^6 M_{\odot}$ model. The primary effect is to stabilize the inner disc, or increase the distance to the region where GI and fragmentation set in.  

 Fragmentation occurs where $Q\le Q_0$ and $t_{\rm cool}<\beta_{\rm crit}/\Omega$. 
With our disc solutions, we calculate where the conditions for fragmentation are satisfied. Following previous simulations that measure fragmentation conditions, we choose $\beta_{\rm crit} = 3$.  
We discuss consequences of the opacity `gap' for where fragmentation occurs in Section~\ref{sec:fragmentation}.

\subsection{Where does fragmentation occur?}
\label{sec:fragmentation}
\begin{figure}
\begin{center}
\includegraphics[width=0.45\textwidth]{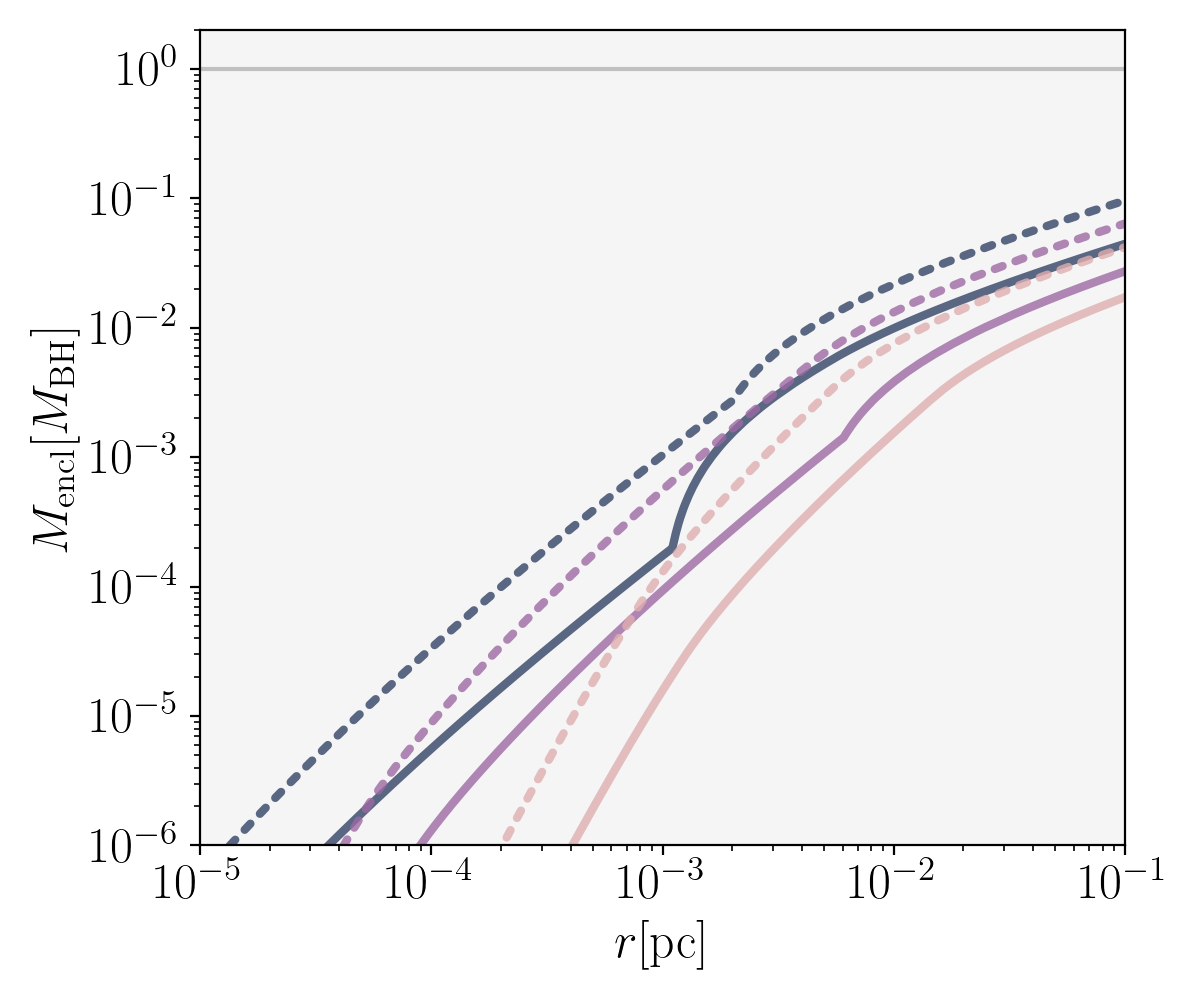}
\caption{Enclosed disc mass divided by the respective central SMBH mass. Same color key as Fig.~\ref{fig:num_discs}.
}
\label{fig:disk_Mencl}
\end{center}
\end{figure}

The fragmenting region of the disc is determined by two concurrent
conditions, namely that $Q\sim Q_0$ and that the cooling time is $t_{\rm cool} \Omega < \beta_{\rm crit}$, where  we choose $Q_0=1.4$ and $\beta_{\rm crit} = 3$.
It is only when both these conditions are satisfied that a disc
can achieve fragmentation. In absence of a short enough cooling time 
the gas temperature will rise as a result of spiral shocks, raising
Q above the fragmentation limit (see \citealt{2007prpl.conf..607D}).
Beyond the radius where this is satisfied (in a regime $\Delta r_{\rm frag}$), the disc will fragment into bound clumps which accrete surrounding material as they collapse, forming protostars. 
Note that the disc does not necessarily need to be more massive than the central SMBH in order to become self-gravitating: if it is thin and dense, the criteria are satisfied before $M_{\rm disc}\sim10^{-2} M_{\rm BH}$ (see Fig.~\ref{fig:disk_Mencl}).

While theory predicts that discs around more massive BHs are more prone to GI 
\citep{Goodman2003,Lodato2007}, we find that discs around less massive SMBHs may fragment at smaller radii, provided the accretion rate is sufficiently high ($f_{\rm Edd}>0.05$). This is not intuitive---despite the lower gas densities, the cooler temperature allows for a different opacity source to dominate and consequently a faster cooling time. 
Note that if you consider the fragmentation radius in terms of the respective Schwarzschild radii of the central SMBH, for which $r_{\rm S}\propto M_{\rm BH}$, then more massive SMBHs fragment at distances of fewer Schwarzschild radii.
However, the lower densities in discs around less massive SMBHs mean that the fragments (and subsequent AGN stars) are born with comparatively smaller masses, as we will show in Fig.~\ref{fig:rf_M}.
The inclusion of surface irradiation increases the distance to the transition radius by a small factor. For the parameter space explored in this work, it does not prevent fragmentation.

\subsubsection{Centi-parsec fragmentation near opacity gaps}

\begin{figure}
\begin{center}
\includegraphics[width=0.45\textwidth]{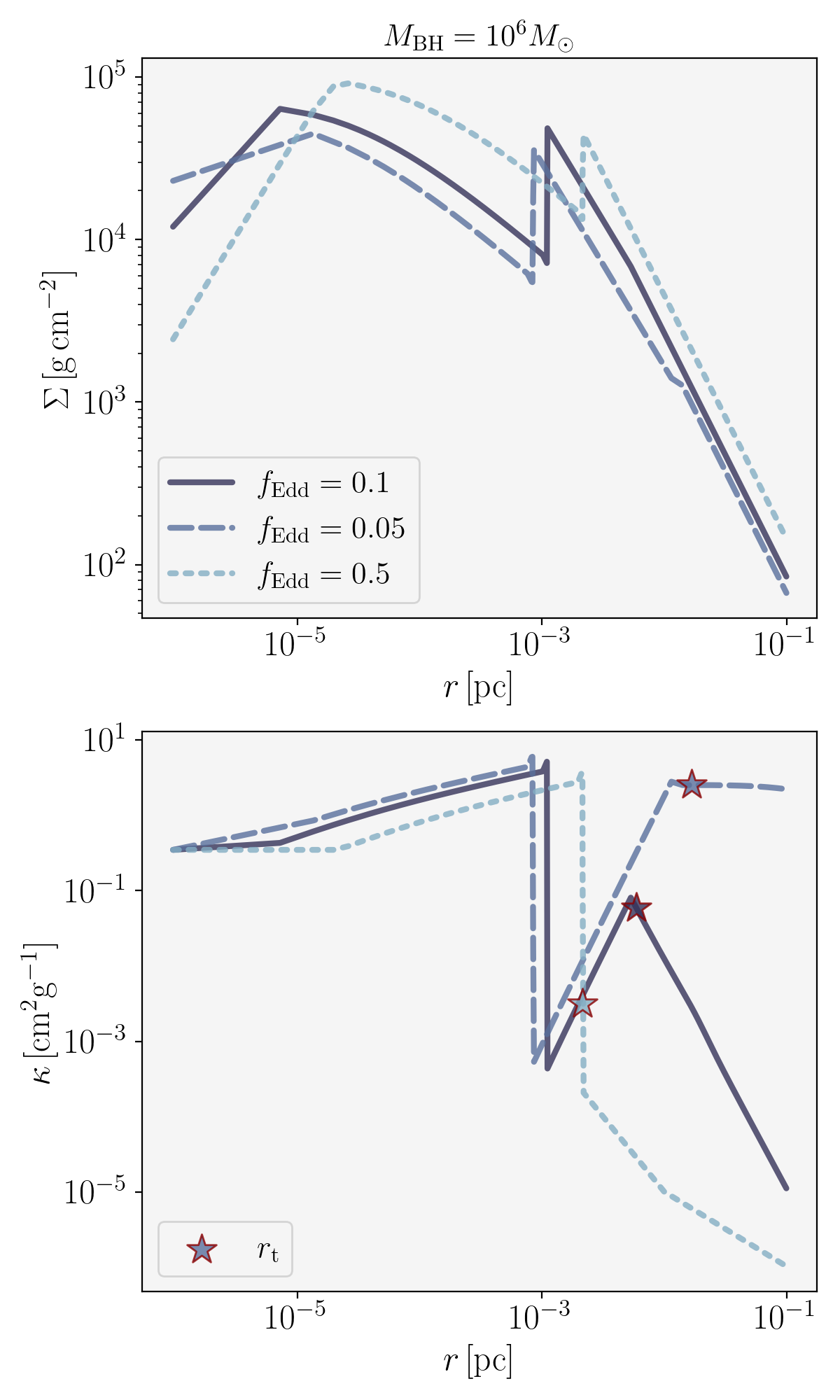}
\caption{Surface density (top panel) and opacity (bottom panel) profiles for $M_{\rm BH} = 10^6 {\rm M_{\sun}}$ and $\alpha = 0.1$ for accretion rates at different fractions of the Eddington rate ($\dot{M}/\dot{M}_{\rm Edd} = f_{\rm Edd}$). Star symbols delineate the transition radius $r_{\rm t}$, or the innermost region where fragmentation occurs. In the inner disc, opacity is dominated by (from left to right) electron scattering, then bound-free transitions until a gap occurs due to $\rm H^-$ absorption, and then molecular lines and dust scattering. Fragmentation occurs within this gap or beyond it, depending on the accretion rate which sets the disc temperature. 
}
\label{fig:disk_fEdds}
\end{center}
\end{figure}

\begin{figure}
\begin{center}
\includegraphics[width=0.49\textwidth]{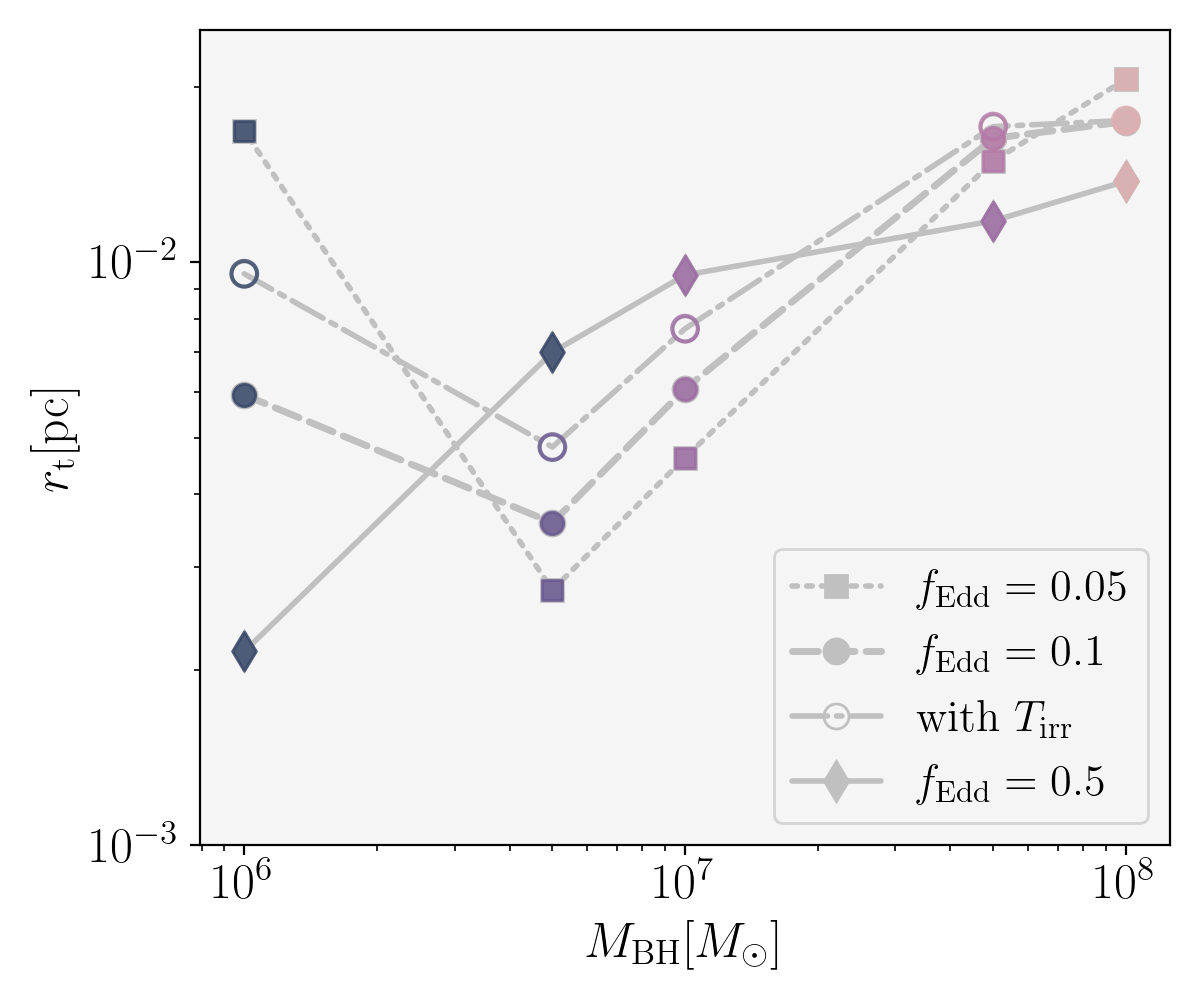}
\caption{Transition radius $r_{\rm t}$ from inner to outer disc (or minimum radius where $Q=Q_0$ and $\Omega t_{\rm cool}<\beta_{\rm crit}$) for discs with $\alpha=0.1$ around a range of SMBH masses. Different lines/symbols correspond to solutions for the corresponding constant accretion rate $\dot{M} = f_{\rm Edd}\dot{M}_{\rm Edd}$.  The hollow circles correspond to the $f_{\rm Edd} = 0.1$ disc with surface irradiation, which pushes fragmentation out to farther radii. Due to opacity transitions at lower temperatures, discs around $10^6 {\rm M_{\sun}}$ SMBHs can fragment at smaller distances depending on the accretion rate.
}
\label{fig:m_rfrag}
\end{center}
\end{figure}

The transition to the fragmenting region of the disc and whether or not this occurs within (or due to) the opacity gap depends on the disc accretion rate.  Fig.~\ref{fig:disk_fEdds} shows solutions for surface density and opacity profiles for discs with $\alpha=0.1$ around a BH of mass $M_{\rm BH}=10^6 {\rm M_{\sun}}$, for three different Eddington ratios. Discs at higher steady-state accretion rates are denser, leading to GI and fragmentation at closer radii ($r_{\rm t}\sim10^{-3}$ pc, denoted by the star symbols). 
The solution with the lowest accretion rate ($f_{\rm Edd}=0.05$) best demonstrates the full opacity gap, which is comparable to solutions found in \citet{TQM2005}. 
A lower accretion rate leads to fragmentation at farther radii ($r_{\rm t}\gtrsim 10^{-2}$ pc) beyond the opacity gap, in a region dominated by dust scattering and sublimation. In the bottom panel of Fig.~\ref{fig:disk_fEdds} we show the opacity profiles for each disc model, where the star symbols indicate beyond which radius the disc is gravitationally unstable. In the inner disc, the opacity is dominated by electron scattering and bound-free+free-free transitions. At intermediate radii, temperatures become low enough for $\rm H^-$ recombination to occur, which results in a sharp opacity drop, beyond which molecular line cooling begins to dominate. At cooler temperatures, the opacity becomes dominated by dust sublimation and dust scattering. 
In these regions the disc is still optically thick, as shown in Fig.~\ref{fig:num_discs}, and thus the transition in opacity strongly affects the cooling rate.
In summary, the dominant opacity law within the fragmentation region may vary from H- absorption to molecular interactions to dust grain sublimation, depending on the combination of density and temperature at $r_{\rm t}$.\footnote{
While we do not address the subsequent implications for opacity changes in this work, we note that the opacity may affect the critical cooling rate \citep{Cossins2010} as well as the resulting stellar properties.}

In Fig.~\ref{fig:m_rfrag} we show the minimum radius of fragmentation, or the transition radius $r_{\rm t}$, versus the central SMBH mass for different accretion rates. Differences in opacity laws determine the cooling rates and lead to the scatter at lower SMBH masses. 
For all solutions shown here, fragmentation begins to occur within $10^{-3}\lesssim r \lesssim 10^{-1}$ pc, but, as we show in the following Section, the variance in disc properties in this region lead to different initial clump properties.

\section{Initial mass distribution of AGN stars}
\label{sec:massesofstars}
In this section we describe how the properties of the disc in the gravitationally unstable region lead to predictions for the size of clump formation, accretion rates onto pre-stellar cores, and an initial mass distribution of AGN stars. 

\subsection{Fragmentation into bound gas clumps}
\label{sec:clumps}

In regions where the disc fragments, gravitationally bound clumps\footnote{We use `clump' and `fragment' interchangeably to refer to the gravitationally bound clouds of gas.} arise with a range of sizes centered around a characteristic scale. In the linear theory, this characteristic scale $\lambda_{\rm mu}$ is derived from the dispersion relation for  local axisymmetric density perturbations, and corresponds to the most unstable wavelength 
\be
\lambda_{\rm mu} = \frac{2 c_s^2}{G \Sigma},
\ee
The maximum scale at which fragmentation can occur
is given by twice the most unstable wavelength.  
Pre-stellar clumps are thus typically born with characteristic initial radii $R_{\rm c} = \frac{1}{2}\lambda_{\rm mu}$, 
and the corresponding enclosed mass\footnote{Note that with the assumption that $Q=Q_0$ for a Keplerian disc, the $M_T$ is similar to the Jeans mass,
$M_{\rm J} = \frac{4}{3} \pi \rho \left(\lambda_{\rm J}/2 \right)^3$
where $\lambda_{\rm J} = (\pi c_s^2 / (G \rho))^{1/2}$, 
this further suggests that the initial clumps formed by GI are already too large to maintain stability, and hence should undergo collapse. However, the Jeans criteria is oversimplified for star formation in the ISM, as it underpredicts the characteristic mass when compared to the observed mass function (e.g. \citealt{Bate2005,LeeHennebelle2019}).} 
 is 
\begin{multline}
\label{eq:Mtoomre}
M_{\rm T} = \pi \Sigma \left(\frac{\lambda_{\rm mu}}{2}\right)^2 = \frac{\pi c_s^4}{ G^2 \Sigma}\\
\approx 2 {\rm M_{\sun}} \left( \frac{Q_{\rm crit}}{1.4} \right)^2 
\left( \frac{1}{\alpha} \right)
\left( \frac{r}{0.01 {\rm pc}} \right)^{3/2}
\left( \frac{M_{\rm BH}}{10^6 {\rm M_{\sun}}} \right)^{1/2}
\left( \frac{f_{\rm Edd}}{0.1} \right)
\end{multline}
where we assume a circular volume element with surface density $\Sigma$, 
and we have plugged in a normalisation corresponding to the outer disc properties for a $10^6 {\rm M_{\sun}}$ SMBH.
As discussed in Section~\ref{sec:planetsstars}, the conventional theory over-predicts the size and mass of clumps when compared to simulations of magnetized or hydrodynamic self-gravitating discs. We include it here as a `maximum estimate' in order to compare to previous work.

An improved estimate that takes into account the fact that, in an  differentially rotating disc,  the instability has a global rather than local character and occurs in non-axisymmetric
conditions, namely along spiral arms  
\citep{Boley2010} finds that the perturbations  are dependent on the density enhancement in the spiral arm, which corresponds to a scale given by $\lambda^{\prime}_{\rm T} = \lambda_{\rm T}/\mach^2$ assuming the flow in the arm behaves as an isothermal shock of Mach number $\mach$.
With this estimate, and further assuming finite thickness across the arm, fragments form with a  characteristic masses of
\begin{multline}
\label{eq:Mfrag}
M_{\rm frag} = 2 \lambda_{\rm mu} \frac{\Sigma c_s}{\Omega f_{\rm g}} \\
\approx 0.3 {\rm M_{\sun}} \left(\frac{Q_{\rm crit}}{1.4}\right) 
\left( \frac{0.1}{\alpha} \right)
\left( \frac{2.5}{f_{\rm g}} \right)
\left( \frac{M_{\rm BH}}{10^6 \rm {\rm M_{\sun}}} \right)^{1/2}
\left( \frac{r}{0.01 {\rm pc}} \right)^{3/2}
\left( \frac{f_{\rm Edd}}{0.1} \right)
\end{multline}
where $f_{\rm g}$ is a shape factor of order unity. We consider $M_{\rm frag}$ a fiducial estimate for the initial mass of prestellar fragments, which notably predicts masses that are an order of magnitude less massive than $M_T$. 

\begin{figure}
\begin{center}
\includegraphics[width=0.49\textwidth]{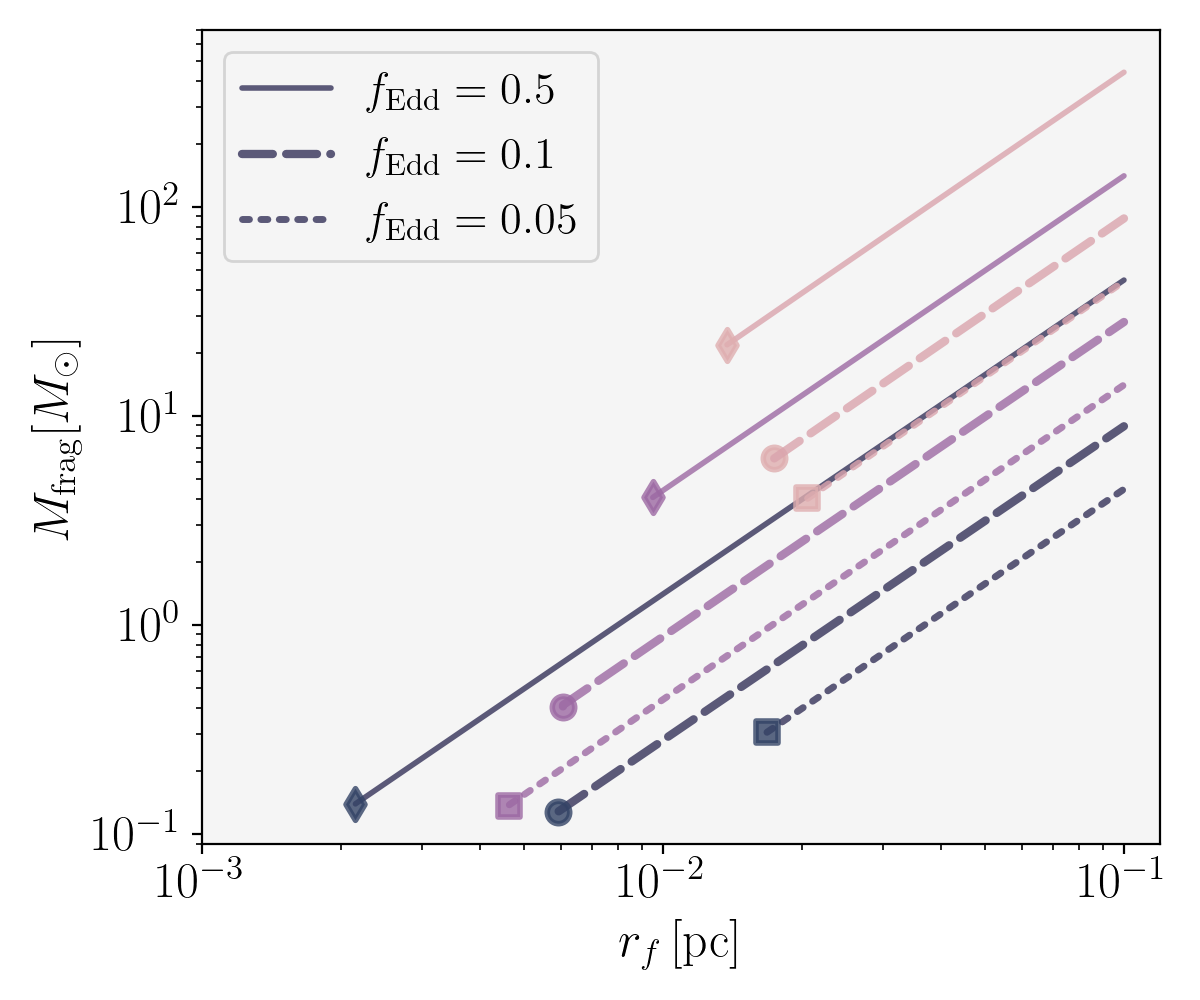}
\caption{The characteristic mass of fragments (Eq.~\ref{eq:Mfrag}) as a function of disc radius in the region where fragmentation conditions are satisfied. The colors correspond to the same scheme as in Fig.~\ref{sec:discmodel}. Discs around higher SMBH masses (lighter lines) produce higher fragment masses. However, this does not correspond to the initial mass distribution of stars, which will depend on the SF efficiency. 
}
\label{fig:rf_M}
\end{center}
\end{figure}

We show the range of clump masses as a function of the fragmentation radius for each disc model in Fig.~\ref{fig:rf_M}. Note that each line represents a value of the characteristic clump mass at each radius, rather than a strict prediction of the mass distribution. A few to several clumps may form in each radial bin with a distribution centered on the characteristic mass. The initial mass distribution of the population will fall within the given mass range, but with a slope that depends on the gas density and SF efficiency.

\begin{figure}
\begin{center}
\includegraphics[width=0.49\textwidth]{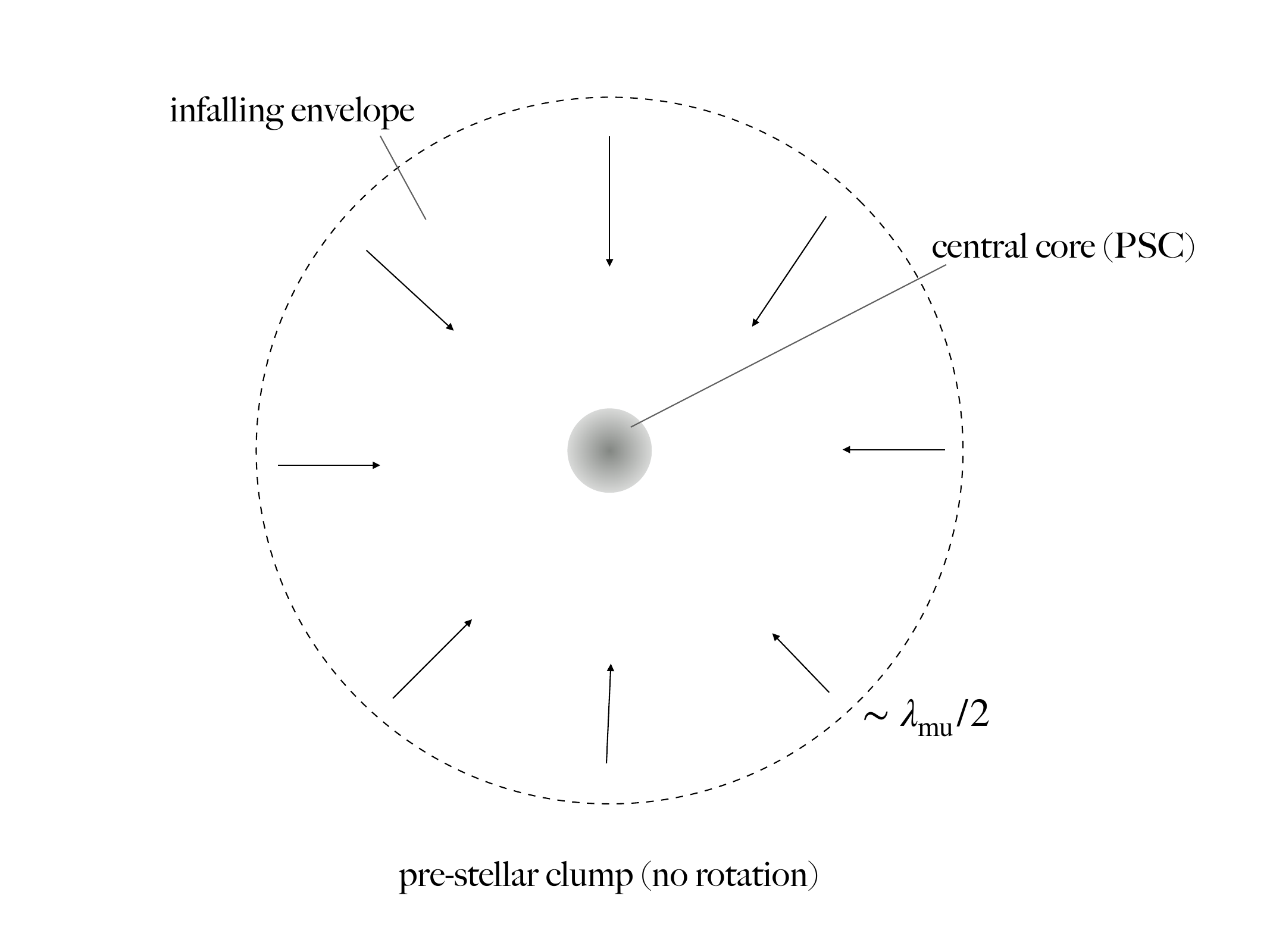}\\
\includegraphics[width=0.49\textwidth]{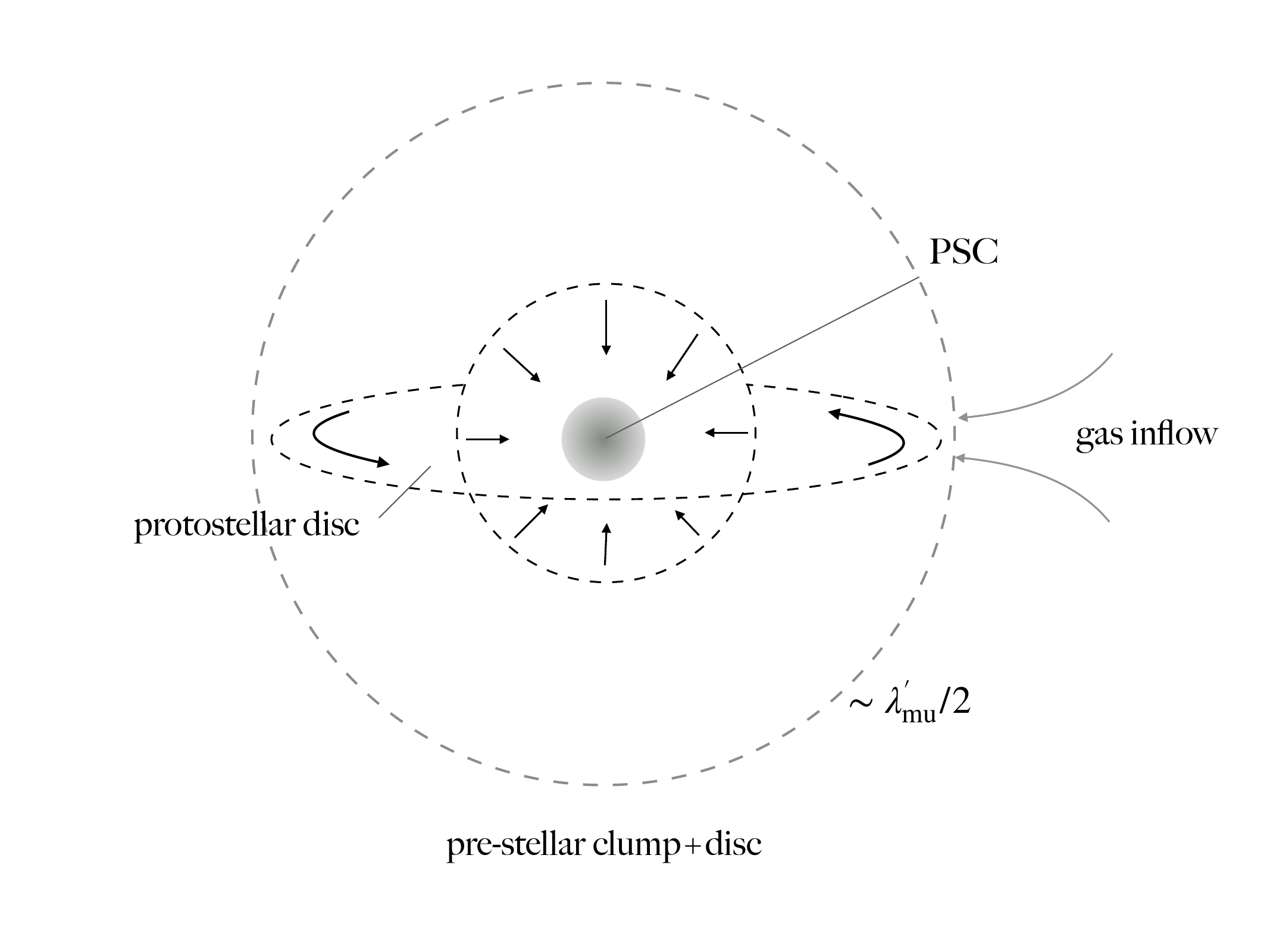}
\caption{Components of the pre-stellar clump  in a non-rotating, isolated case (top) and a disc-embedded, rotationally-supported case (bottom). The clump is a gravitationally bound, collapsing cloud. Neglecting rotational support, the protostar embryo consists of a central hydrostatic core and a surrounding free-falling envelope. For the rotationally-dominated case, the mass of the clump is partly in a central core (the nascent protostar) and partly in a surrounding disc. Additional gas can be supplied to the circumclump disc via gas inflow into the Hill sphere. Accounting for rotational support results in initially less massive stars. }
\label{fig:protostar}
\end{center}
\end{figure}

\subsection{From clumps to protostars}
Once bound, pre-stellar clumps may continue to collapse depending on the gas properties. 
If the central temperature of a fragment is below the Hydrogen dissociation temperature ($\sim2000$ K), collapse of a fragment is initially quasi-static and slow \citep{Decampli1979}. 
Once molecular Hydrogen dissociates, the collapse of fragments occurs dynamically \citep{Larson1969}. 
The former contraction phase is a critical step in conventional star formation (as well as for planet formation in protoplanetary discs, \citealt{Helled2014}), during which the first core forms. In this phase, the surrounding gas is vulnerable to internal feedback once the cloud is optically thick, which slows down the infall of surrounding gas. 
In the AGN case, however, star formation occurs in an environment with initially high temperatures (see the midplane temperature in Fig.~\ref{fig:num_discs}\footnote{
In principle, AGN stars may form in gas that is below the H dissociation temperature, which can occur in discs around less massive SMBHs (in fact, the $10^6 {\rm M_{\sun}}$, $\alpha=0.1$ disc model in Fig.~\ref{fig:num_discs} shows that $T_{\rm mid}(r_{\rm frag}) < 2000 \rm\, K$). In such cases the collapse may be initially delayed but should quickly presume dynamical collapse.}),  
and critically, \emph{the bound clouds are born in a regime where thermal effects from their contraction are unimportant}. 
Instead protostars are effectively born in the main accretion phase, which is analogous to the dynamical collapse phase in planet formation. However, 
we stress that this outcome is where the analogy between fragmentation in protoplanetary discs and in AGN discs breaks down \textemdash in the planet case, protoplanets are vulnerable to disruption during their initially slow contraction \citep{Muller2018}, while in the AGN case, the initially high temperatures and densities ensure that clumps collapse dynamically to form stars and are unimpeded by internal feedback or subsequent disruption.

During (spherical) dynamical collapse, a gravitationally bound clump will collapse on the free-fall timescale: 
\be
t_{\rm ff} = \left(\frac{3 \pi}{32 G \rho} \right)^{1/2}
\approx 10 \left(\frac{r}{0.01 \rm pc} \right)^{} \left(\frac{10^6 \rm {\rm M_{\sun}}}{M_{\rm BH}} \right)^{1/2} \rm yr,
\ee
where $\rho$ is the average density of the cloud, and for the right hand side we have plugged in the disc density scaling from Eq.~\ref{eq:rho}, which provides an upper limit of the free-fall time (since the protostellar clump is ab initio denser than the disc). 
 The estimate is derived for the time for a cloud to collapse to infinite central density, but inevitably before this point the inner temperatures will become sufficiently high to ignite nuclear fusion.  
 The rapid free-fall timescale, a result of the high densities in the disc model, suggests that clumps in the fragmentation region collapse swiftly into AGN stars within  $10^0 \lesssim t_{\rm ff} \lesssim 10^3 \rm \, yr$, depending on the initial density of the gas.  
An additional consequence of such rapid collapse is that fragments become relatively compact, reducing their interaction cross section soon after formation. This suggests that it is unlikely for fragments to merge with each other during this stage of formation, unless a relatively large number of fragments form.

\subsubsection{Accretion during protostar collapse}
During the main accretion phase, protostars continue to grow via accretion from their surrounding envelopes. To provide context on the scale of rapid growth, we describe two scenarios for this process: the first a simplified, inside-out spherical collapse (considered a maximum estimate) and the second is an axisymmetric collapse
in which the growth of the prestellar core (PSC) is governed by the rate of angular momentum
dissipation through a surrounding accretion disc. The latter case further supports the assumptions we adopt in this work, in which the initial clump masses are lower than previously anticipated. 
In Fig.~\ref{fig:protostar} we provide a schematic of the pre-stellar clump assuming purely spherical infall (top panel) and including a rotationally supported accretion disc (bottom panel). 
The latter is 
a natural expectation from fragmentation in an accretion disc. In a shearing medium, PSCs are inevitably born with  angular momentum. Higher angular momentum material settles into a protostellar disc, which can delay collapse and mediate the growth of the core \citep{Galvagni2012}.

In the spherically symmetric case, a PSC accretes from an envelope at a rate governed by the local sound speed, also known as `inside-out collapse' \citep{StahlerPalla2004}: 
\be
\dot{M}_{\rm iso} \approx \frac{c_s^3}{G} = 10^{-2}  {{\rm M_{\sun}} yr^{-1}} 
 \left(\frac{Q_{0}}{1.4}\right) \left(\frac{f_{\rm Edd}}{0.1} \right)
\left(\frac{M_{\rm BH}}{10^6 {\rm M_{\sun}}} \right)
\left(\frac{0.1}{\alpha} \right)
\ee
where $c_s$ is the sound speed of the gas in the co-orbital region and on the right side we have plugged in the parameter scaling implied by $Q=Q_0$, assuming the central BH accretes at a constant fraction of the Eddington rate. This expression assumes the envelope is isothermal and non-rotating, but the infall rate has a weak dependence on rotation \citep{StahlerPalla2004}. 
Given the high temperatures of the AGN disc, inside-out collapse predicts extremely rapid accretion onto protostars ($\dot{M} \gtrsim 0.1-10 \rm \, {\rm M_{\sun}} \, yr^{-1}$, depending on $M_{\rm BH}$). However, this accretion phase is short-lived and only occurs within the sphere of influence of the core, until the nearby gas is quickly depleted.

The growth of prestellar cores that have significant rotation proceeds differently, yet may lead to similarly high accretion rates. In this case accretion occurs via a circumstellar disc, which regulates the flow onto the protostar at a rate set by viscous dissipation (the details of which we neglect here) as well as the gas inflow from the surrounding environment. 
If we consider the shear flow around the protostar, accretion into the Hill radius occurs at a rate (following \citealt{Boley2010})
\begin{multline}
\label{eq:mdothill}
\dot{M}_{\rm Hill} \approx 9\times 10^{-3}  {{\rm M_{\sun}} yr^{-1}} \left(\frac{a}{0.01 {\rm pc}} \right)^{1/2}
\left(\frac{M_*}{\rm M_{\sun}} \right)^{2/3}
\left(\frac{M_{\rm BH}}{10^6 \rm M_{\sun}} \right)^{-1/6}\\
\times \left( \frac{\Sigma}{10^4 {\rm \, g \, cm^{-2}}}\right)
\end{multline}
where we have assumed that the surface density is constant within the protostar region and normalised to values of interest to our system. Not necessarily all of this mass is accreted by the star, given that gas can flow in and out of the Hill radius. Furthermore, accretion via a disc must be regulated by viscous dissipation, which is often inefficient \citep{Shabram2013}, so Eq.~\ref{eq:mdothill} should be interpreted as an upper limit. 

For both the spherical and axisymmetric cases, accretion rates onto collapsing cores are determined by gas properties which, as a consequence of our disc model and the scaling of $M_{\rm frag}$, are constant for all nascent stars.

 For perspective, note that these accretion prescriptions can exceed the Eddington rate for an object of mass $M_*$,
 \be
 \dot{M}_{\rm Edd} = \frac{4 \pi G M_*}{\epsilon_{\rm eff} \kappa_{\rm es} c} 
 \approx 2\times 10^{-8}  {\rm M_{\sun} yr}^{-1} \left(\frac{M}{\rm M_{\sun}}\right) 
 \left(\frac{0.1}{\epsilon_{\rm eff}} \right).
 \ee
Note that we use the electron scattering opacity here, but in cooler regions of the disc where dust can condense, the lower opacity (as shown in Fig.~\ref{fig:disk_fEdds}) leads to a correspondingly higher possible Eddington rate. 

The accretion rate during this phase is high but short-lived, given the rapid free-fall timescale (indeed a strong limit also arises from the local gas supply which quickly runs out). Protostars only increase their masses by at most a factor of $\sim1/10$th during collapse (assuming they accrete over a timescale $t_{\rm ff}$),
resulting in characteristic stellar masses within the range determined by the initial clump mass (Eq.\ref{eq:Mfrag}), as shown in Fig.\ref{fig:m_rfrag}.\footnote{
In the case of SMSs, the accretion rate in the main accretion phase is important for determining the final stellar mass \citep{Hosokawa2013} and the stability of the growing core \citep{Haemmerle2019}. By comparing to core accretion rates of models by \citet{Haemmerle2019}, one can see that even these rapid growth rates remain in the regime of `hydrostatic growth,' i.e. the structure of the PSC can readjust as it gains mass and avoid runaway growth that leads to instabilities.}
In principle, gas inflow from the global accretion disc could continue to feed the circumstellar disc after star formation. At this point we assume that once the star is formed, feedback (either from accretion or stellar winds) would suppress further growth, but we discuss a possible outcome of subsequent stellar accretion in Section~\ref{sec:bondi}.

\subsubsection{Opacity-limited mass}

A minimum clump mass limit can be calculated by determining the point at which the clump becomes opaque to its own radiation. This minimum mass is known as the `opacity limit' mass \citep{Low1976,Rees1976,Silk1977}. Assuming spherical collapse, the opacity-limit mass is of order $M_{\rm min} = 0.025 (1/\mu)^{16/7} (\kappa_f/\kappa_{\rm ES})^{1/7} \, \rm M_{\sun}$, where $\mu$ is the mean molecular weight and $\kappa_f$ the final opacity at which the fragment becomes opaque. The limit is weakly dependent on the opacity and remains below $M_{\rm min}\lesssim 10^{-1} {\rm M_{\sun}}$ for $\kappa_f\lesssim 1 \rm cm^2 g^{-1}$, which is consistent with the range of fragment masses predicted by Eqs.~\ref{eq:Mtoomre} and ~\ref{eq:Mfrag}.

We note that this limit may change if one considers that fragmentation occurs in the vicinity of an accreting SMBH which can impart a temperature background (if the fragmenting material is optically thin). In the presence of a radiation field, the opacity-limited mass becomes more sensitive to the opacity under the assumption that it cannot cool below the background temperature \citep{Low1976}. We defer this calculation to future work, noting that the precise opacity-limit mass in the AGN regime is a complex radiative transfer problem that depends on the composition of the surrounding material, the adiabatic evolution of the fragment, and the preferential geometry of the emitted feedback, from both the AGN and the collapsing fragments.

\subsection{From disc properties to the stellar mass distribution}
\label{sec:stellarbudget}
While the mass distribution shown in Fig.~\ref{fig:rf_M} demonstrates the range of possible stellar masses during a fragmentation episode, it does not quantify the total number of stars formed. For this we must assume a SF efficiency $\epsilon$, which 
for simplicity we assume is constant with radius. 
We adopt $\epsilon= 1\%$ (or $\epsilon=0.01$), which is motivated by efficiencies calculated in models of starburst discs \citep{TQM2005}\footnote{We refer to their disc model with constant gas fraction, which finds an efficiency of $\sim0.02$ in the inner disc region. 
Alternatively, one could self-consistently solve for the reduction in accretion rate as a function of the gas fraction, as is also done in \citet{TQM2005}. In this model a minimum SF rate is required to maintain $Q\sim 1$, which in low opacity regions implies a higher SF rate. } and supported by observations of starburst galaxies \citep{Kennicutt1998} and giant molecular clouds (GMCs) \cite{Murray2011}. 
(We caution, however, that the star formation considered in our model occurs on smaller spatial scales, and need not necessarily lead to similar efficiencies as observed in GMCs.)
Similarly, the efficiency of planet formation via GI in protoplanetary disc simulations suggests $\lesssim 10\%$ of the gas goes into bound clumps, though the fraction is lower if one accounts for subsequent disruption or additional nonlinear effects \citep{Galvagni2012,Muller2018,2021A&A...645A..43S}. 
To remain conservative (and to ensure that star formation does not shut off the accretion flow), we adopt the fiducial value of $\epsilon=0.01$. 
To account for the possibility that star formation is more efficient, e.g. in the case that feedback from star formation is not sufficient to restabilize the disc, we also consider the case of $\epsilon=0.3$, motivated by the maximum star formation rates observed in GMCs. This choice has implications for  the final EMRI rate, which we discuss in Section~\ref{sec:outcomes}.

Stars form over a timescale proportional to the orbital time: $\tau_{\rm ff}\sim \Omega^{-1}$, so the efficiency provides a star formation rate
\be
\dot{M}_* = 2 \pi \int_{r_{\rm t}}^{r_{\rm out}} \dot{\Sigma}_* r dr,
\ee
 where $\dot{\Sigma}_* = \epsilon \Sigma \Omega$ is the star formation rate per unit area, integrated over the fragmenting region from $r_{\rm t}$ to the outer disc edge $r_{\rm out}$. The efficiency can be interpreted as a reduction in the timescale of star formation, or, equivalently, $\epsilon$ tells us what fraction of the gas turns into stars. 
 If no additional gas is supplied to the disc, fragmentation will result in a population of $N_*$ stars
 \be
 \label{eq:nstar}
 N_* = \int_{r_{\rm t}}^{r_{\rm out}}\frac{dm}{dr}\frac{dn}{dm} dr
 \ee
 where $dm/dr$ is the combined mass of stars formed within a radial bin $dr$, which is dependent on the gas density: 
\be
\label{eq:dmdr}
\frac{dm}{dr} = 2 \pi \epsilon r \Sigma(r)
\ee 
The differential mass distribution $dn/dm$ also depends on radius. In each radial bin, $dn$ stars form at masses sampled from a distribution: 
\be
dn = f(m,r) dm
\ee 
where $f(m,r)$ is the radially-dependent mass spectrum for which we choose a normal distribution
\be
f(m,r) = \frac{1}{\sqrt{2\pi}}\frac{1}{\sigma(r)} \exp{\left[\frac{-(m-\mu)^2}{2 \sigma(r)^2}\right]},
\ee
where $\sigma(r)$ is the standard deviation, taken to be $\sigma=M_{\rm frag}(r)/3$ (to ensure that the masses do not exceed twice the characteristic mass), and $\mu = M_{\rm frag}(r)$ is the distribution mean. 

This results in a combined mass of stars 
\be
M_{*,\rm tot} = \int_{r_{\rm t}}^{r_{\rm out}} \frac{dn}{dr} \frac{dm}{dn} dr 
= 2 \pi \epsilon \int_{r_{\rm t}}^{r_{\rm out}} r \Sigma(r) dr,
\ee
and the radial number distribution is constrained by the mass budget as follows:
\be
\frac{dn}{dr} = \frac{d M_{*,\rm tot}}{dr} \frac{dn}{dm}.
\ee
Stars in each radial bin $dn/dr$ follow a mass distribution $dn/dm$ centered on the characteristic mass $M_{\rm frag}(r)$. 
In practice, the distribution is obtained by splitting the radial profiles shown in Fig.~\ref{fig:m_rfrag} into $50$ discrete, logarithmically-spaced bins, and sampling $dn\sim dm/M_{\rm frag}$ stars from a distribution centered on the characteristic mass $M_{\rm frag}(r)$ within each bin. The final distribution does not depend strongly on the number of bins, as long as enough gas is contained within each to form a finite number of stars.  We assume stars form only as single systems, neglecting the binary fraction. 

\begin{figure*}
\begin{center}
\includegraphics[width=0.49\textwidth]{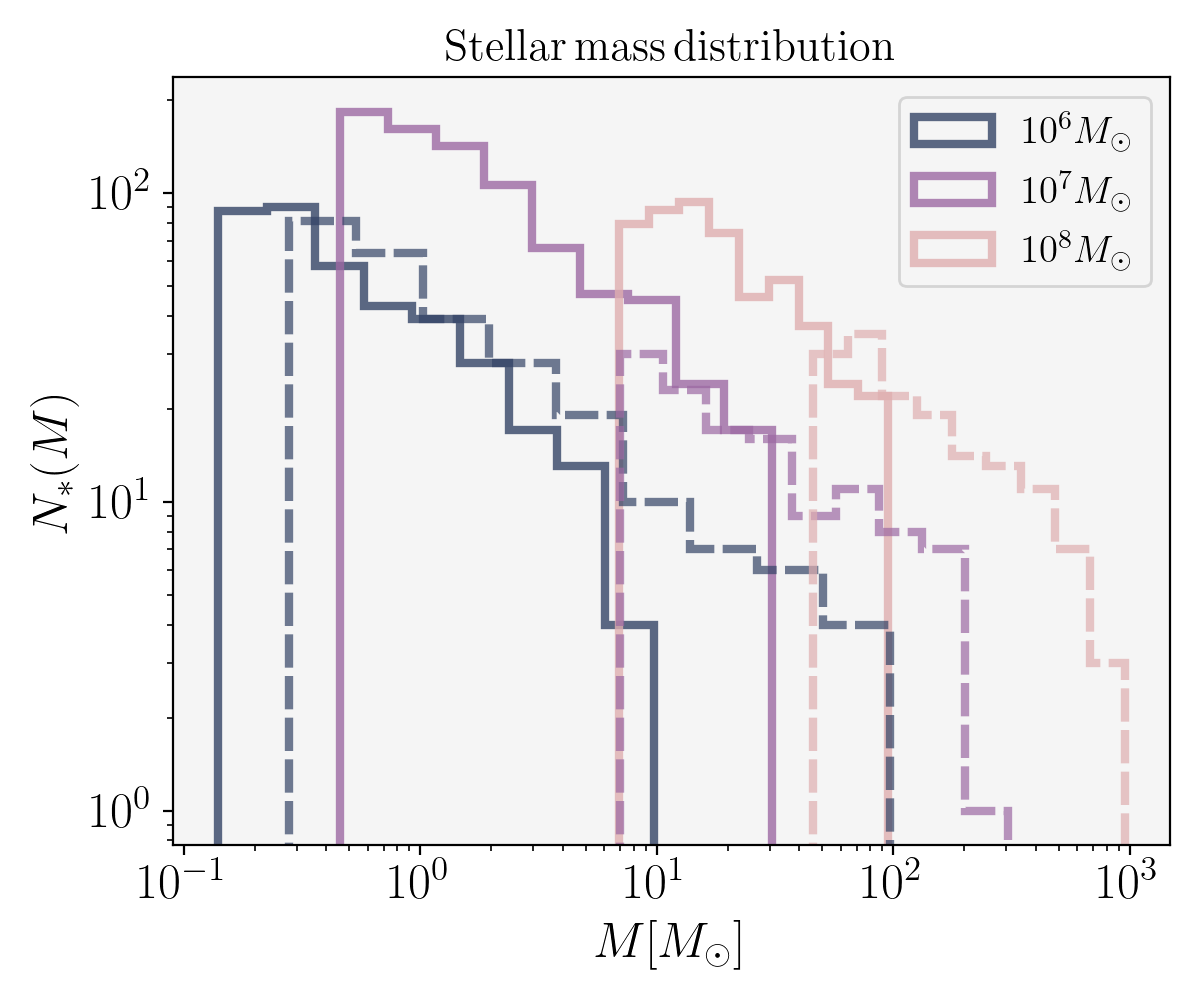}
\includegraphics[width=0.49\textwidth]{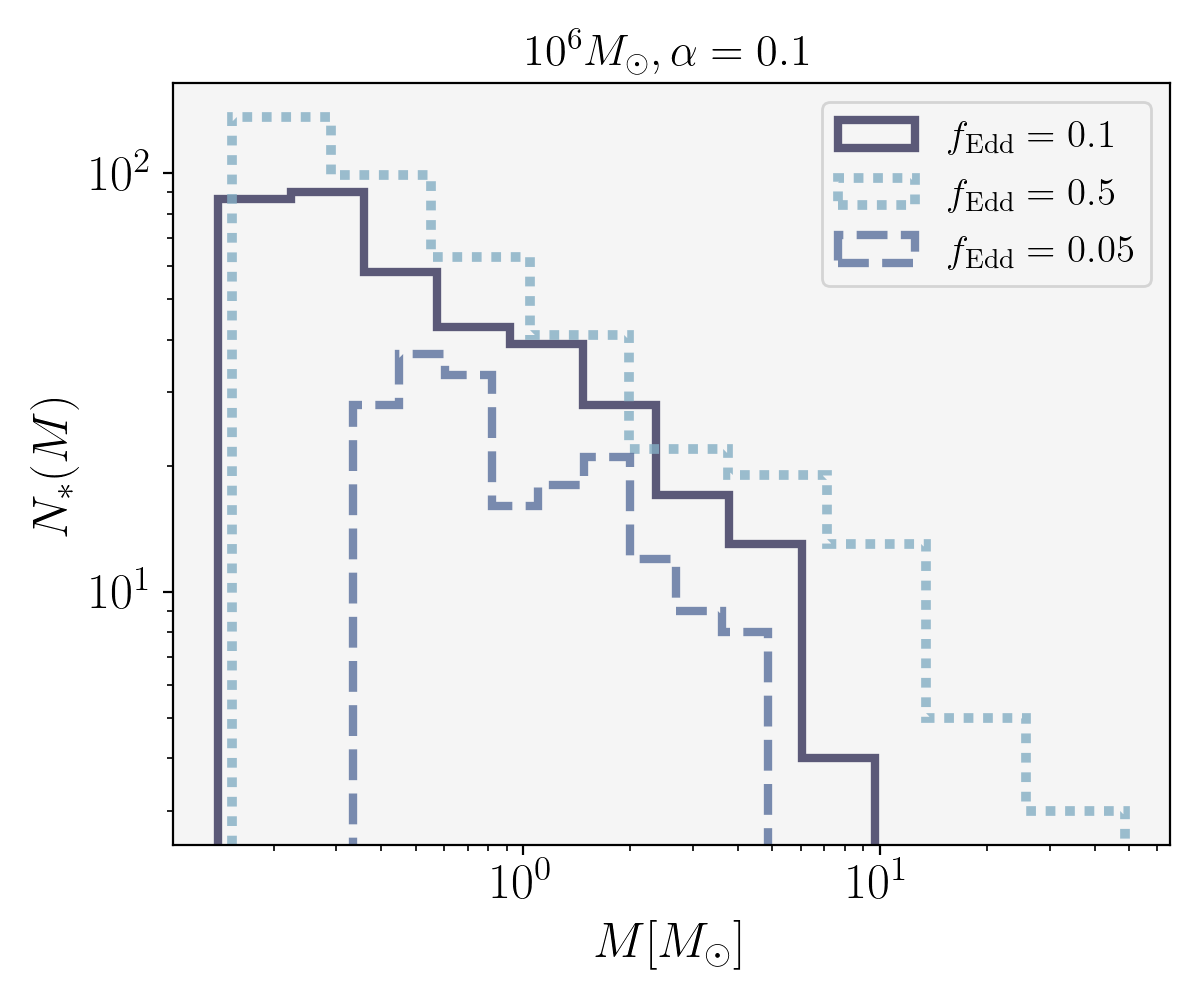}
\caption{ Initial mass distributions of in-situ stellar populations, assuming a SF efficiency $\epsilon=0.01$. Left panel: different SMBH masses with $\alpha=0.1$ (solid) and $\alpha=0.01$ (dashed). Right panel: different accretion rates in units of the Eddington rate, where $f_{\rm Edd} = \dot{M}_{\rm BH}/\dot{M}_{\rm Edd}$. }
\label{fig:Massdist}
\end{center}
\end{figure*}

 This results in the initial mass function shown in Fig.~\ref{fig:Massdist}, for which we plot models with $\epsilon=0.01$. 
 The AGN stellar mass distribution is top-heavy compared to the conventional Salpeter initial mass function (IMF): Fitting a line to the data shown in Fig.~\ref{fig:Massdist} provides a slope  $dN/dm \propto m^{\Gamma}$, with $\Gamma \sim - 0.7$. 
 This is considerably flatter than the Salpeter IMF, where the slope is $\Gamma_{\rm SP}=-2.35$.
 Our finding is in agreement with the stellar mass distributions measured in simulations by  \citet{Mapelli2012} of a fragmenting disc for a Sgr A* type SMBH. For different gas masses and thermodynamic assumptions, they measure an IMF slope ranging from $\Gamma \sim -0.5$ to $-1.6$. 
In another related set of simulations by \cite{NayakshinCuadraSpringel2007}, fragmentation of a disc around Sgr A* also produces a top-heavy IMF that is dependent on the stars subsequent accretion of the surrounding gas.
Such a top-heavy IMF is also supported by observations of ubiquitous metal enrichment in AGN, which can be produced by supernovae of massive stars \citep{Wang2011,Toyouchi2022,XuBian2018,LaiFuyan2022,DittmannJermynCantiello2022}.

Our choice of a constant SF efficiency (e.g. Eq~\ref{eq:dmdr}) requires that the number of stars formed decreases at outer radii to ensure that the gas disc is not depleted. 
The consequence of this is that the star forming region of the disc is finite, and only a small number of stars form at outer radii, where the heaviest end of the distribution lies. 
Note that in the case of a higher SF efficiency, the amplitude of the distribution increases according to Eq.~\ref{eq:dmdr}, while the the range and slope of the distribution remains the same.
In principle the disk could extend to larger radii, in which case a higher SF efficiency would potentially increasing the heavy end of the distribution and the resulting EMRI rate. 
To keep the model conservative, we choose to set the outer disk edge to be $r_{\rm out} = 0.1$ pc, which is also
 in agreement with constraints on disk sizes from observed spectral energy distributions (see \citealt{SirkoGoodman2003}). 
In the case that the efficiency varies with radius, the slope of the IMF will change (as will the steady-state $\dot{M}$ profile of the disc). 
Furthermore, we neglect here the possibility that fragments may merge, an increasing possibility for cases of higher $\epsilon$, which would skew the distribution to higher masses.

\subsection{Bondi accretion onto embedded stars}
\label{sec:bondi}
Here we demonstrate that despite the lower masses of in-situ stars, accretion from their environment at limited rates can rapidly skew the mass distribution to higher values. 
For stars moving on prograde, circular orbits (such that the relative velocity between the orbiter and the gas is negligible), the Bondi-Hoyle accretion rate is given by \citep{Hoyle1939,BondiHoyle1944,2004NewAR..48..843E}
\be
\dot{M}_{\rm Bondi} = \pi r_{\rm B}^2 \rho c_s
\ee
where $r_{\rm B} = 2 G M_*/c_s^2$ is the accretion radius, which is defined by the region of gravitational influence. This estimate suggests rapid accretion onto embedded stars, and has been presented as a rapid growth mechanism for gas-embedded stellar populations \citep{Davies2020}. In fact, the disc properties derived above suggest a Bondi accretion rate that exceeds the accretion rate onto the central SMBH, which would break more than the steady state disc assumption. 
However, the Bondi radius is in most cases larger than the disc thickness, and thus the assumption of rapid, spherically symmetric infall is not appropriate, especially if stars grow to larger masses. 
While accretion inside the Bondi radius may occur in a super-Eddington configuration\textemdash  certain accretion/feedback geometries allow for super-Eddington growth, e.g. a slim disc \citep{Abramowicz1988} or accretion with non-equatorial feedback \citep{Jiang2014} or reduced radiative efficiency \citep{Jiang2019}\textemdash 
the rate must at least be limited by the amount of gas that can flow into (and stay within) the star's sphere of influence. 

To demonstrate an outcome of stellar accretion, we apply a modified Bondi rate to the initial population for which the accretion radius is limited to either the star's Hill radius or the disc thickness, such that
\be
r_{\rm acc} = \frac{1}{2}{\rm minimum}(r_{\rm Hill}, h(r_*))
\ee
where $r_{\rm Hill} = r_*({M_*}/{3 M_{\rm BH}})^{1/3} $, $h(r_*)$ is the disc scale height at the stars position $r_*$, and the factor of 1/2 ensures that the accretion radius is well within the disc\footnote{ See also \citet{Dittmann2021} for a more detailed discussion of stellar outcomes via disc-influenced Bondi accretion. Note however, that their stellar evolution models do not explore the high densities as found in our disc models.}
(and consequently so that the rate remains at least an order of magnitude below the disc accretion rate). The accretion rate is given by 
\be
\label{eq:mdotstar}
\dot{M}_{\rm *} = \eta \pi r_{\rm acc}^2 \rho c_s,
\ee 
where we introduce an efficiency factor of $\eta = 0.01$ to account for empirical evidence that gas flowing into the sphere of influence does not guarantee its accretion onto the star. In fact, simulations show that gas flows past an embedded migrator from the outer to the inner disc, (see e.g. \citealt{Duffell2014} or \citealt{Derdzinski2019}).  

\begin{figure}
\begin{center}
\includegraphics[width=0.49\textwidth]{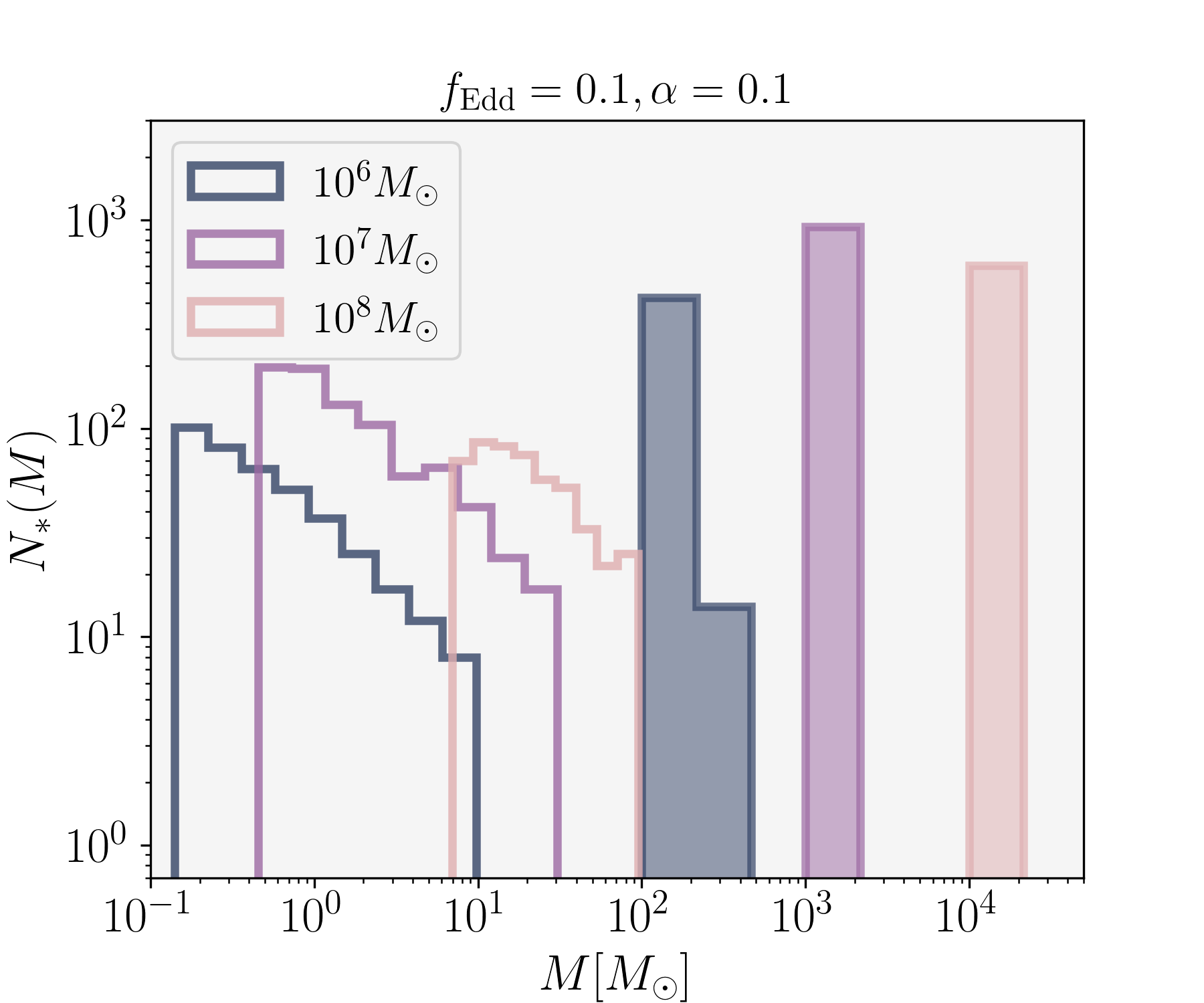}\\
\includegraphics[width=0.49\textwidth]{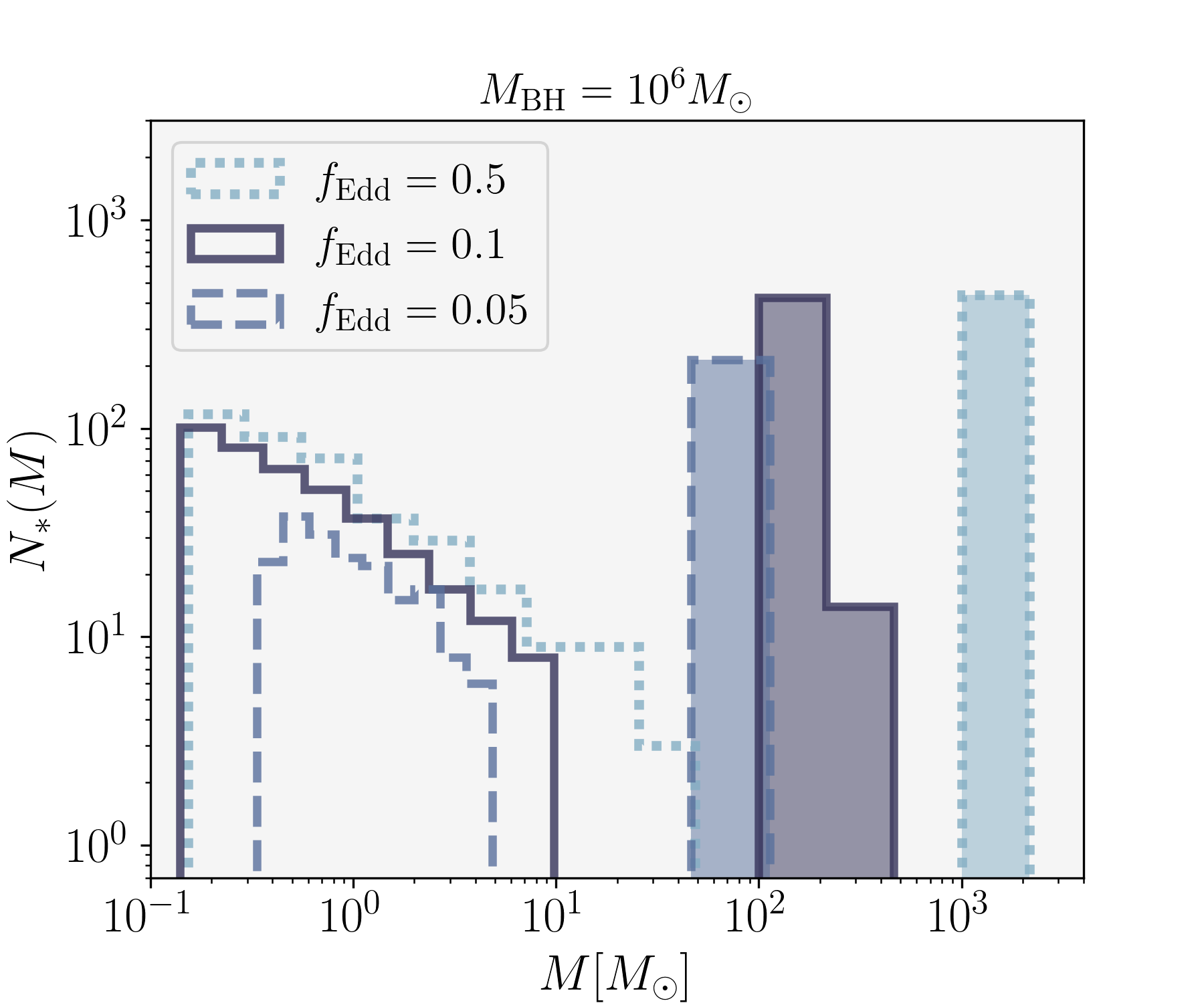}
\caption{Same as Fig.~\ref{fig:Massdist}, for $\alpha=0.1$. Shaded histograms show the mass distribution after stars accrete via the (limited) Bondi-rate prescription for $\tau_{\rm AGN}\sim10^7$ yr. }
\label{fig:Massdistgrowth}
\end{center}
\end{figure}

Fig.~\ref{fig:Massdistgrowth} shows the evolution of the stellar mass distribution after $\sim 10^7 \,\rm yr$, 
a timescale set by the AGN disc lifetime discussed in Section~\ref{sec:evolution}. 
Here the accretion rates are determined by the disc properties at the initial position and updated as the stellar mass increases. 
This assumes the lowest disc densities, leading to a conservative estimate of the Bondi rate, but 
also the highest disc thickness, which allows for a larger accretion radius. This accretion increases stellar masses by more than two orders of magnitude throughout the disc lifetime. The result is more pronounced for discs with $\alpha=0.01$, for which the higher disc density leads to more rapid growth (in these cases we limit the stellar masses to $10^4 M_{\sun}$). The shift in peak of the stellar mass distributions for all models is shown in Table~\ref{table:rates}.

We consider this accretion rate an overestimate, given that during growth, stars will likely reach a critical mass beyond which accretion becomes inefficient due to radiative feedback or momentum-driven winds. The precise limit depends on details of stellar winds, feedback, and the accretion geometry, but may lie between $30$ to $60 {\rm M_{\sun}}$ (\citealt{Ekstrom2012, Cantiello2020}, although see \citealt{DittmannJermynCantiello2022}). 
Due to the uncertainties regarding stellar accretion, in the following section we use the initial mass distribution (shown in Fig.~\ref{fig:Massdist}) as our fiducial population, which is based on the fragment mass limits from Eq.~\ref{eq:Mfrag}. 
As we show, including stellar accretion will alter the final numbers when determining observational outcomes, but the basic story remains the same: stars of different masses will have different fates due to the interplay of migration, stellar evolution, and tidal disruption.

\section{Subsequent evolution and migration of AGN stars}
\label{sec:evolution}
\begin{figure*}
\begin{center}
\includegraphics[width=1.\textwidth]{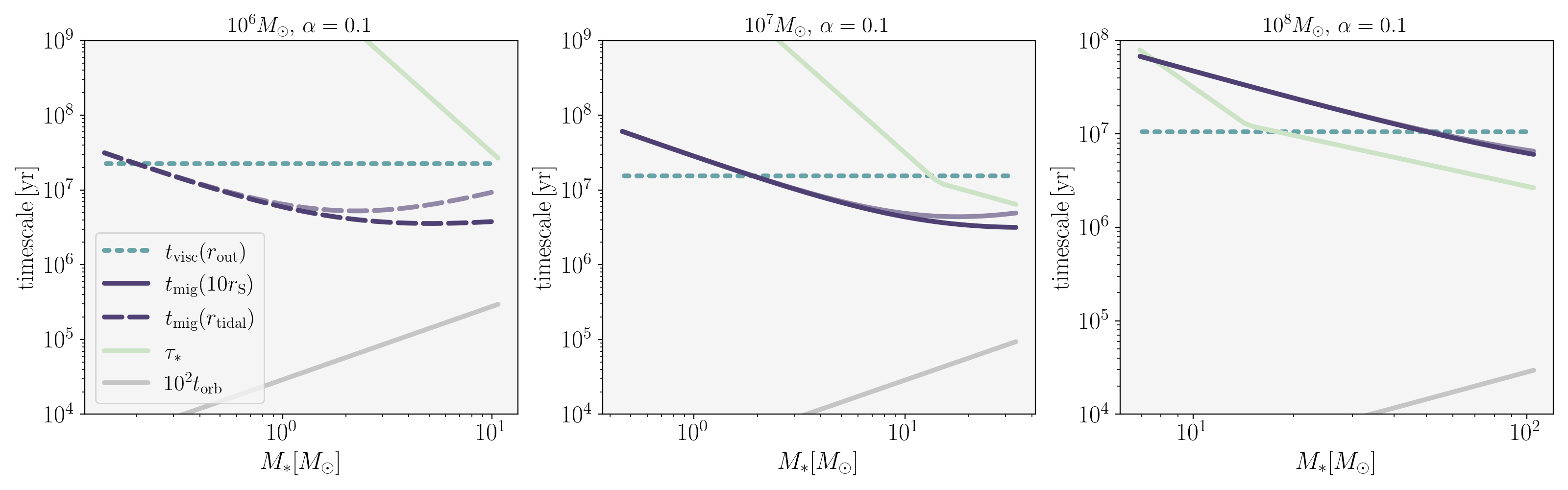}\\
\includegraphics[width=1.\textwidth]{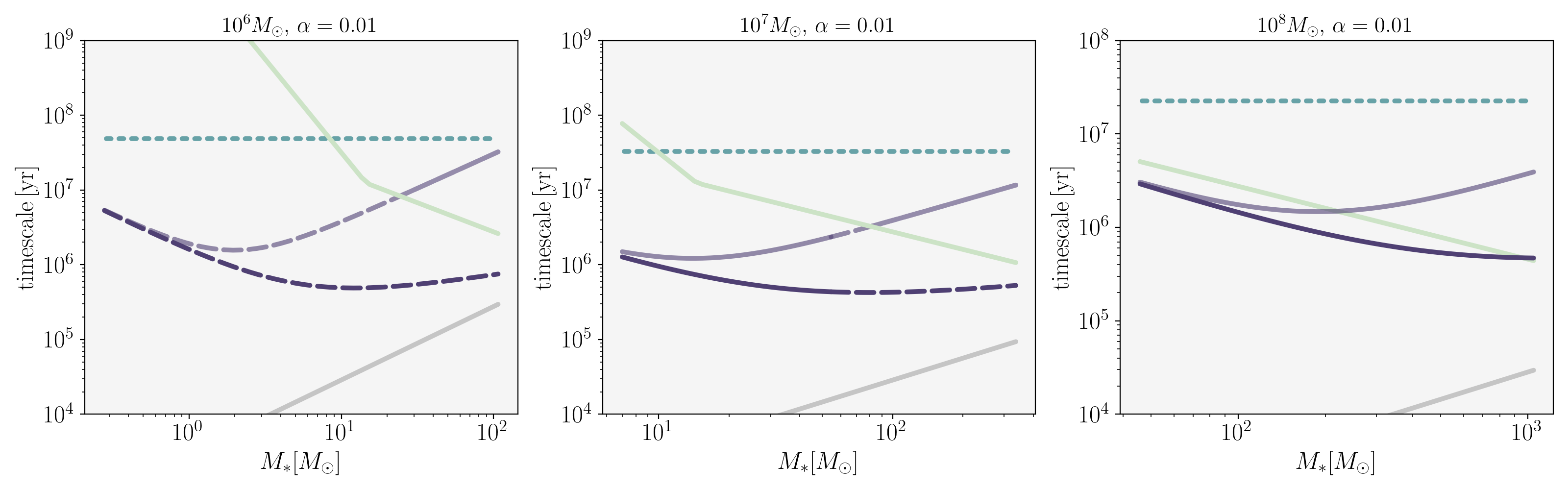}
\caption{Migration timescale from intial separation to $10 r_S$ (solid purple lines), or to $r_{\rm tidal}$ (dashed purple lines), stellar evolution timescale (green line), AGN lifetime (blue dashed line), and orbital time (grey line) for three different SMBH masses and two disc viscosities. Migration times are shown for Type I (dark purple line) and with a gap-opening correction (faded purple line) which significantly slows down migration for higher stellar masses. 
Stars in the left panels are disrupted before they leave compact remnants, while higher mass stars in the middle and rightmost panels survive disruption to be accreted, producing in-situ EMRIs.}
\label{fig:timescales}
\end{center}
\end{figure*}

We now turn to the subsequent migration of AGN stars.   PSCs collapse on a timescale faster than migration (see the dynamical time in Fig.~\ref{fig:timescales}), which allows us to calculate the subsequent evolution of AGN stars separately from their formation.
We break down the range of outcomes for AGN stars into three categories: 
\begin{enumerate}
    \item Tidal `disruption': Stars migrate inwards until they reach the tidal radius where they are destroyed. 
    \item In-situ EMRIs: Stars migrate inwards until they are accreted by the SMBH, either by avoiding disruption (depending on the relationship between the SMBH mass and stellar structure) or collapsing to compact remnants. 
    \item Leftover stellar population: Stars migrate inward until the disc disperses, surviving the AGN phase.
\end{enumerate}
Each of these outcomes is dependent on the mass of the star, the mass of the SMBH, and the disc properties. For the present work we neglect details that would occur as sub-categories (e.g. whether an EMRI occurs as a star or a specific stellar remnant, it is categorized as outcome (ii)). 
Our fiducial model assumes that no further accretion occurs after star formation, meaning that we adopt \emph{initial} mass distributions in Fig.~\ref{fig:Massdist} and keep the stellar masses constant.  We discuss implications of including stellar accretion (as shown in Sec.~\ref{sec:bondi}) in Sec.~\ref{sec:discussion}.

Stars embedded in the disc will excite nonaxisymmetric perturbations that react on the star's orbit, resulting in an exchange of angular momentum that will affect the orbital separation \citep{1980ApJ...241..425G}. In the limit of small mass ratio (here neglecting dependencies on gas thermodynamics), this migration regime is referred to as `Type I,' and the strength of the (typically inward) torque on an embedded migrator of mass $M_*$ at a separation $r$ in the isothermal limit can be estimated with the following expression \citep{Tanaka2002}:
\be
\label{eq:typeI}
T_{\rm I} = - \frac{1}{2f_{\rm I}} \Sigma(r) r^4 \Omega^2 \left(\frac{M_*}{M_{\rm BH}}\right)^2 \left(\frac{h}{r}\right)^{-2}
\ee
where $f_{\rm I}$ is a factor that depends on disc gradients. Rather than solving for $r_{\rm I}$ from our disc solutions which can lead to sign changes in the torque, here we fix $f_{\rm I}=2$ to represent a typical inward migration rate. In Appendix~\ref{sec:traps}, we discuss the implications of disc gradients for migration reversal.
This corresponds to a change in radial separation $\dot{r}_{\rm I} = 2 T_{\rm I} / v_{\phi} / M_*$, where $v_{\phi} = (G M_{\rm BH}/r)^{1/2}$ is the Keplerian orbital velocity.

In the classical paradigm, migration slows down if the migrator carves a gap, reducing the density in the coorbital region, which weakens the migration torque. In our case, gap-opening for AGN stars is unlikely for stars with small  mass ratios compared to the central BH (e.g. for $q\lesssim 10^{-5}$), as suggested by conventional gap-opening criteria \citep{Crida2006}. Furthermore, a number of numerical studies conducted for
self-gravitating discs on various scales have shown that migration in such systems can be even faster than implied by Eq.~\ref{eq:typeI} due to the contribution of torques from asymmetric flows in the vicinity of the perturber \citep{Malik2015,2013CQGra..30x4008M}, which either inhibits
gap formation completely  or at least suppress the formation of a deep gap that can slow down migration sensibly \citep{SzulagyiMayer2017}.
We note that the migration of a \emph{population} of stars may not be well-represented by expressions that assume a single migrator. 
Simulations tend to show that while a 
fraction of fragments tend to migrate inward at a rate comparable to the Type I rate, other fragments may experience migration that slows down or reverses direction \citep{Helled2014}. 
In the inner disc, gap-opening or partial gap-opening may occur, especially if stars are able to accrete and grow. 
For simplicity we consider only inward migration but with two limiting estimates (Type I and a slower, gap-corrected expression).

 In the case that a gap begins to open, numerical simulations find in many cases that the torque on a perturber scales with the reduced surface density in the gap, and can be estimated by \citep{2014ApJ...782...88F,2018ApJ...861..140K}
\be
\label{eq:tgap}
T_{\rm gap} = \frac{1}{1+0.04K} T_{\rm I},
\ee
where $K = q^2 (r/h)^5 \alpha^{-1}$. This estimate takes into account the possibility of partial gap-opening, allowing for a continuous expression for the torque as a function of $q$ and disc parameters. 
When calculating migration rates for the stellar population, we take into account the reduction in gas density due to star formation. This only results in a minor effect for models with high SF efficiency ($\epsilon=0.3$), for which the migration times increase by a small factor $\lesssim 2$.

We caution that the above expression assumes a non-self gravitating, isothermal accretion disc, and is not tested for all disc properties or thermodynamics. Furthermore, it is known that gas dynamics in the gap-opening regime become highly nonlinear, and the resulting torque can have nontrivial dependence on disc parameters \citep{Duffell2014}, especially when resolving gas close to the perturber \citep{Derdzinski2019} or when gaps are particularly steep \citep{ChenZhang2020}.  However, due to the majority of stellar masses being below the gap-opening threshold, the correction has minor implications for the outcome of AGN stars (discussed in Section~\ref{sec:outcomes}).

Additionally, orbital angular momentum loss occurs due to GW radiation from interaction with the central SMBH, which, assuming a circular inspiral, results in a separation evolution \citep{Peters64}
\be
\dot{r}_{\rm GW} = -\frac{64}{5} \frac{G^3}{c^5} \frac{(M_{\rm BH} + M_*)^3}{(1+1/q)(1+q)} \frac{1}{r^3}
\ee
where $G$ is the gravitational constant, and $c$ is the speed of light.

The timescale for a star to migrate from its initial separation $r_{\rm i}$ to a final separation $r_{\rm f}$ is
\be
t_{\rm mig} = \int_{r_{\rm i}}^{r_{\rm f}} \frac{1}{\dot{r}_{\rm I}+\dot{r}_{\rm GW}} dr 
\ee
assuming that gas-induced migration and GW migration are uncoupled. 
More massive stars migrate faster and reach the inner disc first, despite being born at farther separations. 
In theory migration should slow down once the star becomes a less massive compact remnant. We neglect this detail here as it will only affect when, not if, the stars produce EMRIs.

Stars will migrate till they are accreted by the central SMBH or disrupted by tidal forces. Thus the outcome of stars relies on the interplay between the migration timescale and the stellar lifetime. Tidal disruption\footnote{
Tidal disruption typically requires that stars reach the tidal radius on an eccentric orbit (which is also a consequence of dynamical formation scenarios in the absence of gas), and in this case the star is violently disrupted. 
In the case of disc-embedded stars with near circular orbits, it is more likely that stars are tidally \emph{dissolved}, or experience Roche lobe overflow \citep{Rossi2021,Dai2013,MetzgerStone2017}, and that torques from the disc may complicate the picture. We can conveniently still refer to this event as a tidal dissolution event (TDE). 
} 
occurs if stars reach the tidal radius, or the region where tidal forces from the central SMBH overcome the gravitational binding energy of an incoming star, defined by
\be
\label{eq:rtidal}
r_{\rm tidal} \approx \left( \frac{M_{\rm BH}}{M_{*}} \right)^{1/3} R_*
\ee
where $R_*$ is the radius of a star with mass $M_*$ \citep{Hills1975}. The expression is approximate due to uncertainties in stellar structure, for which we estimate stellar radii by
\be
R_* = \left(\frac{M_*}{{\rm M_{\sun}}} \right)^{1/2} \rm R_{\sun}
\ee
which underestimates the size of sub-solar mass stars\footnote{
Sub-stellar objects or brown dwarfs may survive disruption, contributing to the `eXtremely small mass ratio inspiral event rate (see Fig. 1 in \citealt{2019PhRvD..99l3025A}}
as well as massive stars, at least in isolated environments (see e.g. \citealt{Hosokawa2013}).
We neglect partial disruptions, which may occur for massive stars with extended envelopes \citep{Rossi2021,Macleod2012} 
and tidal heating, which may inflate stellar radii \citep{AlexanderMorris2003,GongjieLoeb2013}. 
Tidal disruption is the primary obstacle for survival of stars around less massive SMBHs ($M_{\rm BH}\lesssim 10^7 {\rm M_{\sun}}$), for which the tidal radius lies outside of the SMBH event horizon. Disruption can be avoided if stars collapse to become CRs before migrating to this regime.

The stellar lifetime also depends on the stellar mass: 
\be
\label{eq:taustellar}
\log \tau_{*} = A \log(M_*/{\rm M_{\sun}}) + B
\ee
 where $A$ follows from the mass luminosity relation ($L_*\propto M_*^{1-A}$). For stars below $ M_*/{\rm M_{\sun}}< 15$, we adopt the main sequence values $A = 2.5$ and $B=10$. For stars more massive than $M_*/{\rm M_{\sun}} \ge 15$, we adopt scalings from rotating massive stellar models at solar metallicity (\citealt{Ekstrom2012}, but see also \citealt{Yusof2013}) for which $A=-0.78$ and $B=8.00$. 

The final timescale for comparison is the AGN disc lifetime, which 
is given by the viscous time at $r_{\rm out}$, assuming that no additional gas is supplied to the disc during the accretion episode:
\be
\tau_{\rm AGN} = t_{\rm visc} = \frac{2}{3}\frac{r^2}{\alpha c_s h} \sim 10^7 {\rm yr} 
\left(\frac{0.1}{\alpha}\right)
\left(\frac{0.02}{h/r}\right)^2
\left(\frac{10^6 {\rm M_{\sun}}}{M_{\rm BH}}\right)^{1/2}
\ee
where we have plugged in the turbulence assumption for kinematic viscosity.

The migration times shown in Fig.~\ref{fig:timescales} assume a constant stellar mass, without taking into account that beyond a stellar lifetime, stars `become' a compact remnant of smaller mass.
This does not change the migration time considerably, since most of the migration occurs during the stellar lifetime. The main implication of the remnant mass is whether the EMRI occurs during the AGN disc lifetime or after disc dispersal. More massive remnants will inspiral more quickly due to stronger GW emission, 
and thus more likely occur while the disc is still present.

In Fig.~\ref{fig:timescales}, we show the timescales for Type I migration (dark purple lines) from the starting separation to either the tidal radii or to $10 r_{\rm S}$, where $r_{\rm S} = 2 G M_{\rm BH}/c^2$ is the Schwarzschild radius, depending on which occurs first, as a function of the stellar mass. For comparison we show the dynamical time at the formation radius (grey line), the stellar lifetime (green line), and the AGN disc lifetime (blue dashed line). Intersections in the lines determine the minimum mass below which stars do not survive to leave CRs or form EMRIs.
Migration is also shown with the gap correction (faded purple lines, using Eq~\ref{eq:tgap}), which slows down migration for stars with mass ratios greater than $q\gtrsim10^{-6}$, suggesting that for each disc model more massive stars survive disruption. 

Gas-driven migration on its own becomes inefficient at inner radii (at $r\lesssim 10^{-5}\rm pc$), where the disc is thick and the gas density is low. However, in this regime GWs become important and soon dominate the inward migration, leading to comparatively fast migration timescales, around $t_{\rm mig} \sim 10^6 - 10^7 \rm yr$ depending on the disc model and stellar/remnant mass. 
In most cases, the migration time to disruption or accretion is less than the stellar lifetime. In the case of disruption, this suggests that stars do not survive to leave CRs or produce in-situ EMRIs.

\subsection{Other migration regimes}
\label{sec:other}

Simulations of Type I migration in the non-isothermal and isothermal limits find more detailed expressions for the torque that depend on the density, temperature, and entropy gradients of the disc \citep{KleyCrida2008,Tanaka2002,Paardekooper2010}.  
With our disc solutions we can calculate the torque profiles and determine for which regions the isothermal or adiabatic torque regimes occur. These calculations and resulting torque profiles are shown in Section~\ref{sec:traps} of the Appendix.
We find that the magnitude of torques is within an order of magnitude of the Type I estimate from Eq.~\ref{eq:typeI}, whether it is in the adiabatic or isothermal regime. 
More interestingly, our disc solutions suggest the presence of regions where the torque changes direction from inward (reducing angular momentum of the migrator) to outward (increasing its angular momentum), typically at distances $\sim10 - 10^2 r_{\rm S}$ from the SMBH. 
The intersection of these regions is referred to as a `migration trap' in the context of both planets \citep{Lyra2010} and BHs
\citep{Bellovary2016}.

If star and/or compact object migration is halted in these regions, this could increase the interaction rate between migrators \citep{Secunda2019}. 
However, it is likely that migration continues due to other processes and that such traps are not sustained (e.g. GW emission). In particular, the highly dynamic regime of self-gravitating discs is unlikely to preserve any feature requiring
a delicate torque balance. 
Motivated by the latter, we neglect migration traps in our calculation of the stellar orbital evolution and resulting EMRI rate. Instead we adopt the assumption that migration is fast, inward, and rapid, following the scaling in Eq.~\ref{eq:typeI}, which is a choice more in line,
qualitatively, with how migration is currently understood in self-gravitating discs. 
Type I migration leads to a conservative in-situ EMRI rate, as the inclusion of the gap correction or traps will delay the migration of stars, providing more time for them to collapse to compact remnants. It is also supported by recurring evidence of fast migration in discs with GI (e.g. \citealt{Malik2015}, but see the code comparison paper by \citealt{Fletcher2019}).

\section{Outcome of the AGN star population}
\label{sec:outcomes}
By comparing the migration timescales to stellar lifetimes and AGN disc lifetimes, we can categorize outcomes of AGN stars. We focus primarily on in-situ EMRIs before briefly touching on TDEs and leftover stellar populations.

\subsection{In-situ gas-embedded EMRIs}
 Under the assumption that no additional gas is fed to the disc, we calculate a {\it minimum EMRI rate} per AGN during an isolated accretion episode. 
 Such a scenario is plausible if accretion onto the SMBH produces feedback that inhibits additional gas from falling into the SMBH Bondi radius. The rate is an underestimate if SMBHs experience multiple episodes of accretion over longer timescales. 

The in-situ EMRI rate depends on the efficiency of star formation (how much of the gas turns into stars within $\Delta r_{\rm frag}$), the stellar mass distribution (determined by disc properties), and subsequent migration rates. 
From one star formation event, we can quantify the total number of stars formed as $N_{*}$ (Eq.~\ref{eq:nstar}).  Whether stars become EMRIs is determined by the timescale comparisons in Fig.~\ref{fig:timescales}. In order to be accreted by the SMBH within the disc lifetime, stars must satisfy three constraints that depend on their masses and initial separations: 
\begin{enumerate}
    \item Stars migrate toward the SMBH with the AGN disc lifetime ($t_{\rm mig}<\tau_{\rm AGN}$), which means that their mass must be above a critical mass $M_{\rm i}$. 
    \item Stars avoid disruption because the tidal radius (Eq.~\ref{eq:rtidal}) is within the regime where they become GW sources ($r_{\rm tidal} < 10 r_{\rm S}(M_{\rm BH})$). This translates to their mass being below a limit $M_{\rm ii}$ (e.g. as shown in Fig.~\ref{fig:timescales} for $M_{\rm BH}=10^7 {\rm M_{\sun}}$).
    \item Stars avoid disruption because they collapse to compact remnants before reaching the tidal radius ($\tau_{*} < t_{\rm mig}$), which requires that their masses are greater than a limit $M_{\rm iii}$ so that their lifetime is more rapid than the migration time. 
\end{enumerate}
In other words, EMRIs originate from stars where the migration timescale (solid purple lines in Fig.~\ref{fig:timescales}) is less than the AGN disc lifetime, excluding regions where stars are disrupted (dashed purple lines). The total number of EMRIs per AGN  is given by summing up the stars from the mass distributions (e.g. Fig.~\ref{fig:Massdist}) within the mass range that satisfies these conditions, such that
$N_{\rm emri} =  N_{*}[(M_*>M_{\rm i})\,  {\rm and} \, ((M_*<M_{\rm ii}) \,{\rm or}\, (M_* > M_{\rm iii} ))]$.

\begin{figure}
\begin{center}
\includegraphics[width=0.49\textwidth]{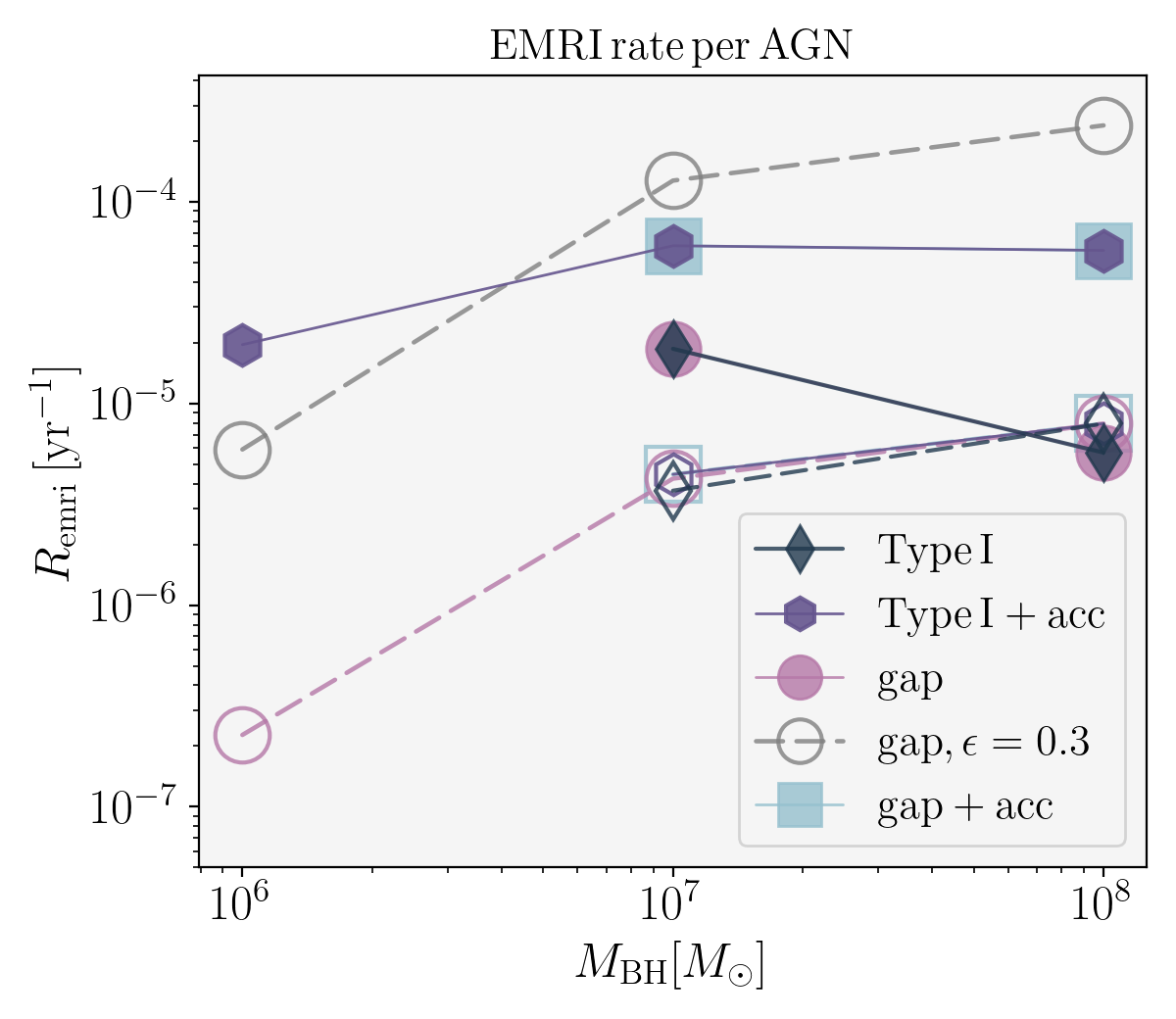}
\caption{EMRI rate per AGN, for discs with $f_{\rm Edd}=0.1$ with a star formation efficiency $\epsilon=0.01$. Solid symbols show rates for discs with viscosity parameter $\alpha=0.1$, hollow symbols for discs with $\alpha=0.01$. We include the EMRI rate for discs models with Type I migration (diamonds), Type I migration plus stellar accretion (hexagons) gap-corrected migration (circles), gap-corrected migration plus stellar accretion (squares). All models 
assume a star formation efficiency $\epsilon=0.01$, except for the gray circles which adopt $\epsilon=0.3$.}

\label{fig:Remris}
\end{center}
\end{figure}

Over an accretion disc lifetime $\tau_{\rm AGN}$, the in-situ EMRI rate per accretion episode is
\be
R_{\rm emri} = N_{\rm emri}/\tau_{\rm AGN}.
\ee
We show this EMRI rate for all disc models in Fig.~\ref{fig:Remris} and in Table~\ref{table:rates}.  Not all models produce in-situ EMRIs. Specifically, the survival of stars in the $10^6 {\rm M_{\sun}}$ BH disc model is sensitive to assumptions of stellar accretion and migration. Stars formed in this system are at lower masses, and efficient (Type I) migration leads to their disruption within $\sim 10^6 \rm yr$. If stars can accrete to larger masses, however, or if migration slows down due to partial gap-opening, this allows for a subset of the population to produce compact remnants. In these cases, the EMRI rate is $R_{\rm emri}\sim 10^{-7}-10^{-5} \rm yr^{-1}$.
For disc models with larger SMBH masses, in-situ EMRIs occur at rates $R_{\rm emri}\sim 10^{-5}-10^{-4} \rm yr^{-1}$ depending on the disc properties, the torque prescription, and the SF efficiency.  
The inclusion of stellar accretion boosts the EMRI rate, in some cases by an order of magnitude. 
The effect of the gap-correction is less clear: in principle it allows for more massive stars to survive disruption and produce EMRIs (this is seen in the bottom middle panel of Fig.~\ref{fig:timescales}). In practice, most stars are below the mass where the gap-correction becomes relevant, as shown in Sec.~\ref{sec:massesofstars}, unless they accrete substantially. 
The higher viscosity disc (with $\alpha=0.1$) for $M_{\rm BH}=10^7 M_{\sun}$ produces more EMRIs, despite the stars having slower migration times. This is due to the lower density discs producing larger numbers of less massive stars that can avoid disruption. 
In a similar fashion, models with high SF efficiency lead to a higher EMRI rate, simply by increasing the number of massive stars that survive to seed EMRIs.

We note that by limiting the disc size to $0.1$ pc, we also limit the number of massive stars that form, and in turn limit the EMRI rate. We expect these rates will increase for larger discs that experience higher SF efficiencies. On the other hand, 
these rates may decrease if one considers mutual gravitational scattering between stars/BHs, which may preserve some stars from disruption but also lose potential EMRIs \citep{ForganRice2018}. Interactions between BHs which can merge with each other via gas-assisted binary formation \citep{Tagawa2020a} could also decrease the rate, while producing higher-mass EMRIs (or IMRIs).

The rates we find are in agreement with the rate of in-situ EMRIs predicted by the models of \citet{DittmannMiller2020}.
For further comparison, \citet{PanYang2021b}, compute EMRI rates due to  interactions of nuclear cluster stellar/BH population with a massive accretion disc. 
For the `conventional' EMRI scenario in dry nuclei (via two-body interactions), they find EMRI rates around $\sim 10^{-6} \rm \, [yr^{-1}]$ per galactic nuclei. In the presence of a gas disc, the EMRI rate typically increases to $\sim 10^{-6} - 10^{-5} \rm \, yr^{-1}$ per AGN, with an order of magnitude variance depending on the mass of the SMBH and the model parameters. These numbers are in line with our calculations for models that produce EMRIs, suggesting that the in-situ channel can generate EMRIs at a comparable rate to the orbit-capture channel.

\citet{Tagawa2020a} consider the evolution of a population of stellar cluster stars and BHs that are captured into binaries in an AGN disc. This work is focused on the evolution of stellar mass binaries via gas-assisted mergers, but also predicts a consequent EMRI rate $0.1-0.6\,\rm Gpc^{-3} yr^{-1}$. For comparison, we can express the EMRI rate in terms of the central SMBH accretion rate: $R_{\rm m} = N_{\rm emri}/\dot{M}_{\rm BH}(\tau_{\rm AGN}) \,\rm M_{\sun}^{-1}$. This can be used to derive a volumetric in-situ EMRI rate $ \tilde{R} = R_{\rm m}\dot{M}_{\rm SMBHs} $, where $\dot{M}_{\rm SMBHs}$ is an estimate of the total accretion onto all SMBHs in the local universe. Calculations by \citet{2004MNRAS.351..169M} suggest $\dot{M}_{\rm SMBHs}\sim5\times10^{-6}  - 5\times 10^{-5}\,{\rm M_{\sun} yr^{-1} Gpc^{-3}}$ for  $z\sim 0$ to $1$, respectively (neglecting variance in $f_{\rm Edd}$ or dependence on SMBH mass).
For our model, the per-AGN EMRI rates for the $10^6 {\rm M_{\sun}}$ SMBH vary from $R_{\rm emri}\sim10^{-7} - 10^{-5} \rm yr^{-1}$ for an SMBH accretion rate $f_{\rm Edd} = 0.1$. Using the above, 
we derive a ($z\sim0$) volumetric rate of $\tilde{R} = R_{\rm emri}\dot{M}_{\rm SMBHs} / \dot{M}_{\rm BH} \sim 0.5 - 10 \, \rm yr^{-1} Gpc^{-3}$, 
though we caution that the true rate will be more dependent on the fraction of AGN that are near-Eddington. 
This simple estimate suggests that the in-situ EMRI rate may be comparable to or greater than the rate derived via AGN-disc assisted binary capture. 

Note that these rates do not necessarily correspond to the detectable rate, given that the more massive SMBH in our disc models produces EMRIs with GW frequencies too low to be detectable by LISA. \emph{Detectable} EMRIs via dynamical processes (e.g. two body encounters) in dry nuclei are expected to occur at a rate $10-10^3 \rm\, yr^{-1}$ according to \citet{2017PhRvD..95j3012B}. 
Overall our results suggests an additional boost to the EMRI rate driven by in-situ formation in sufficiently massive accretion discs.

\begin{table*}
\begin{tabular}{cccc|ccccc}
 SMBH mass & Viscosity & Accretion rate &  frag. radius 
 & Stellar growth & Mass dist. peak &Torque   & EMRI rate ($\epsilon=0.01$) & ($\epsilon=0.3$) \\
 $M_{\rm BH}$ [$\rm M_{\sun}$] &
{  $\alpha$ } &
{  $f_{\rm Edd}$ } &
{$r_{\rm t}$ [\rm pc] } &
{$\dot{M}_*$} &
{ M[{\rm max}(N_*($M$))] [$\rm M_{\sun}$] } &
{} &
{ $R_{\rm emri} [\rm yr^{-1}]$}   &
{ $R_{\rm emri} [\rm yr^{-1}]$}
\\
\hline
$10^6$ & $0.1$ & $0.1$ & $0.006$ 
& \makecell{ none \\ none \\ Bondi$^*$ \\Bondi$^*$ }  
& \makecell{ 0.2 \\ 0.2 \\ 215\\215} &  
\makecell{ Type I \\ gap \\Type I \\ gap } 
& \makecell{ $0$ \\ 0 \\$1.95\times10^{-5}$ \\0 }  
& \makecell{ 0 \\ 0 \\ {} \\ {} } \\
   \specialrule{0.2pt}{1pt}{1pt}

$10^6$ & $0.01$ & $0.1$ & $0.002$ 
& \makecell{ none \\ none \\ Bondi$^*$ \\Bondi$^*$ } 
& \makecell{ 0.5 \\ 0.5 \\ $10^4$ \\ $10^4$ } &  
\makecell{ Type I \\ gap \\Type I \\ gap } & 
\makecell{ 0 \\ $2.27\times10^{-7}$ \\0 \\ 0 } &
 \makecell{ 0 \\ $5.9\times10^{-6}$ \\ {} \\ {} }\\
   \specialrule{0.2pt}{1pt}{1pt}

$10^6$ & $0.1$ & $0.05$ & $0.017$ 
& \makecell{ none \\ none \\ Bondi$^*$ \\Bondi$^*$ }  & 
\makecell{ 0.6 \\ 0.6 \\ 113 \\  113 } &  
\makecell{ Type I \\ gap \\Type I \\ gap } & 
\makecell{ 0 \\ 0 \\0 \\ $5.97\times10^{-6}$ }    & 
\makecell{ {} \\ {} \\ {} \\ {} }\\
\specialrule{0.2pt}{1pt}{1pt}
 
$10^6$ & $0.1$ & $0.5$ & $0.002$ & \makecell{ none \\ none \\ Bondi$^*$ \\Bondi$^*$ }  & 
\makecell{ 0.3 \\ 0.3 \\ 2150 \\ 2150 } &  
\makecell{ Type I \\ gap \\Type I \\ gap } & 
\makecell{ 0 \\ 0 \\0 \\ 0 }   & 
\makecell{ - \\ - \\ - \\ - } \\
   \specialrule{0.2pt}{1pt}{1pt}   
    
 $10^7$ & $0.1$ & $0.1$ & $0.006$ & \makecell{ none \\ none \\ Bondi$^*$ \\Bondi$^*$ }  & 
 \makecell{ 0.7 \\ 0.7 \\ 2150 \\ 2150} &  
\makecell{ Type I \\ gap \\Type I \\ gap } & 
\makecell{ $1.86\times10^{-5}$ \\ $1.86\times10^{-5}$ \\$6.04\times10^{-5}$ \\$6.04\times10^{-5}$ } & 
\makecell{ $5.64\times10^{-4}$ \\ $5.64\times10^{-4}$ \\ {} \\ {} } \\
   \specialrule{0.2pt}{1pt}{1pt}   
    
 $10^7$ & $0.01$ & $0.1$ & $0.008$ & \makecell{ none \\ none \\ Bondi$^*$ \\Bondi$^*$ }  & 
 \makecell{ $16 $ \\$16 $ \\ $>10^4$\\ $>10^4$} &  
\makecell{ Type I \\ gap \\Type I \\ gap } & 
\makecell{ $3.69\times10^{-6}$ \\ $4.23\times10^{-6}$ \\$4.45\times10^{-6}$ \\$4.45\times10^{-6}$ } & 
\makecell{ $1.11\times10^{-4}$ \\ $1.27\times10^{-4}$ \\ {} \\ {} }  \\
   \specialrule{0.2pt}{1pt}{1pt}
 $10^8$ & $0.1$ & $0.1$ & $0.017$ & \makecell{ none \\ none \\ Bondi$^*$ \\Bondi$^*$ }  & 
 \makecell{ 12 \\ 12 \\ $>10^4$ \\ $>10^4$} &  
\makecell{ Type I \\ gap \\Type I \\ gap } & 
\makecell{ $5.70\times10^{-6}$ \\ $5.70\times10^{-6}$ \\$5.73\times10^{-5}$ \\$5.73\times10^{-5}$ }  & 
\makecell{ $1.75\times10^{-4}$ \\ $1.75\times10^{-4}$ \\ {} \\ {} }  \\
   \specialrule{0.2pt}{1pt}{1pt}   
 $10^8$ & $0.01$ & $0.1$ & $0.013$ & \makecell{ none \\ none \\ Bondi$^*$ \\Bondi$^*$ }  & 
 \makecell{ $90$ \\$90$ \\ $>10^4$\\ $>10^4$} &  
\makecell{ Type I \\ gap \\Type I \\ gap } & 
\makecell{ $7.94\times10^{-6}$ \\ $7.94\times10^{-6}$ \\$7.94\times10^{-6}$ \\$7.94\times10^{-6}$ }  & 
\makecell{ $2.39\times10^{-4}$ \\ $2.39\times10^{-4}$ \\ {} \\ {} }  \\
   \specialrule{0.2pt}{1pt}{1pt}
\end{tabular}
\caption{System parameters, stellar population characteristics, and EMRI rates for each subsequent disc and migration model. From left to right we show the central SMBH mass, the disc viscosity parameter, the SMBH accretion rate as a fraction of the Eddington rate, the radius of transition $r_{\rm t}$ from the inner stable disc to the fragmenting region, the stellar accretion model (either none or limited Bondi), the peak of the stellar mass distribution, the migration model (Type I or with a gap correction), and finally, the resulting in-situ EMRI rate assuming a star formation efficiency $\epsilon=0.01$ or $\epsilon=0.3$. We disregard models with stellar accretion for high $\epsilon$, as this would lead to a large population of stars depleting the gas reservoir.   }
\label{table:rates}
\end{table*}

\subsection{Final distribution of stars and compact remnants }

While most stars are disrupted or accreted by the the central SMBH, 
a portion of the population can be left over after disc dispersal if the migration timescale is longer than the disc lifetime. This occurs for lower mass stars in discs  with higher viscosity parameter $\alpha=0.1$ (or lower densities) for which the accretion lifetime $t_{\rm AGN}\sim10^7 \rm\, yr$. 
The maximum mass of these remnant populations scales with the mass of the the central SMBH.
For the central SMBH masses of $\{10^6,10^7,10^8 \} {\rm M_{\sun}}$ SMBH, stars with masses below $\{0.1, 2, 50 \} {\rm M_{\sun}}$ survive the disc phase, respectively. 
These ranges may change if the stars accrete substantially during their evolution, since an increase in mass will affect the migration rate.  
Accounting for more complex migration scenarios, such as the possibility
of migration traps or, even more likely, the mutual gravitational interactions between AGN stars concurrently with disc-driven migration, would
diminish the importance of conditions at birth for the stellar population.
Furthermore, given that we only consider a single AGN activity cycle, one should be careful not to compare our results directly to observed stellar populations (e.g. stellar discs around Sgr A*) around SMBHs that may have a more complex accretion history.

\subsection{Tidal disruption}
\label{sec:tdes}
 When determining the in-situ EMRI rate, we exclude `disrupted' stars since the outcome of disruption depends on the stellar structure. 
It is likely that rather than being promptly disrupted, these stars experience Roche Lobe overflow \citep{MetzgerStone2017} which may or may not be stable, depending on the impact of the accretion disc.
 For the $10^6 {\rm M_{\sun}}$ SMBH discs in particular, surviving tidal forces from the central SMBH is the main obstacle for in-situ stellar populations. 
 These stars might be relevant for the production of Quasi-periodic eruptions \citep{MetzgerStoneGilbaum2022} or some version of TDEs\footnote{Depending on the mass loss rate of the star and its interior structure, the stellar orbit may evolve adiabatically as its internal density readjusts \citep{Rossi2021}, allowing stars to recede from the SMBH. However, the torques from global gas disc will alter this evolution.}, which we plan to investigate in a follow-up paper.
If these stars survive partial disruptions, they may continue to migrate into the inner disc and potentially increase the in-situ EMRI rate.

\section{Discussion}
\label{sec:discussion}

In this work we show that gravitational instability in radiatively efficient, massive accretion discs produces unique stellar populations.  Subsequent evolution of embedded stars can lead to three outcomes: tidal destruction of stars, accretion of stars (or their remnants) as extreme mass ratio inspirals, or leftover stars after disc dispesral. The outcome primarily depends on the interplay of the stellar lifetime and migration timescales,  which are dependent on the disc structure. 
Of course, not all AGN are the same\textemdash accretion rates and geometry can vary, and our fiducial model is one of many.  We demonstrate that within a single disc model, observables such as the in-situ EMRI rate are dependent on the host SMBH and accretion rate. Our disc model is appropriate for massive discs with near-Eddington inflow rates, or quasars, and not low-luminosity, radiatively inefficient accretion flows which are more common in the local Universe. A proper volumetric rate calculation should take into account the quasar population as a function of SMBH mass and redshift.

Our work complements previous studies on in-situ stellar populations in AGN accretion discs which also suggest that star formation in AGN discs should produce stellar populations with large masses \citep{GoodmanTan2004,Levin2007}. One critical difference in our work is that we neglect the interaction between in-situ stars, which may merge with each other. We hypothesize that the realistic outcome is likely a combination of the two. The interaction between stars could reduce the EMRI rate, yet lead to more massive EMRIs.  Determining at what rate stars merge or evolve independently will require a thorough investigation of the mutual interactions within the population. 

In another sense our model is conservative, given that we adopt a relatively low star formation efficiency for the environment considered ($\epsilon=1\%$) as well as corrections to the initial fragment masses. Both of these ingredients act to reduce the mass of the stellar population. 
Even under these constraints, our models predict that each accretion episode can produce a combined stellar mass of $\sim10^2-10^4 {\rm M_{\sun}}$ during a single isolated accretion episode. SMBHs may experience several accretion episodes, or active duty cycles, during their cosmological evolution \citep{2016MNRAS.458..816H}, which will have implications for the cosmological EMRI rate.\\

Several recent works consider how stellar evolution may vary when embedded in a denser, hotter environment than the typical ISM \citep{Cantiello2020,Dittmann2021,Jermyn2021,2022ApJ...929..133J}. These models predict a range of evolutionary outcomes depending on the initial distance of the star to the central SMBH, the accretion mechanism, and chemical mixing. Their choice of initially solar mass stars is consistent with the range of initial stellar masses derived in this work for discs around $10^6 {\rm M_{\sun}}$ SMBHs, however the density in our disc models is considerably higher (by at least $3$ or more orders of magnitude), which is beyond the capability of the numerical stellar evolution codes employed in those works. 
How stellar evolution is altered in higher density environments (along with aspherical accretion or radiative feedback) remains to be explored.
The expected outcome based on the models by \citeauthor{Cantiello2020} is that stars will accrete rapidly but also eject stellar winds. This ultimately limits their growth to $\sim\!10{\rm s} - 100{\rm s} \, \rm M_{\sun}$. The constant supply of fuel can lead to a population of `immortal stars', particularly in the inner disc region \citep{2022ApJ...929..133J}. We expect that incorporating this aspect into this work will affect the population of stars that survive tidal disruption by increasing their lifetime, though in reality this will depend on nuances of accretion and feedback. 
Overall these works highlight the importance of understanding the interplay between stellar evolution, accretion, feedback, and migration. 
\\

A strong prediction of gravitational instability is an initial increase in mass inflow due to the generation of structures that increase angular momentum transport. This is supported at different scales: One example at stellar scales arises from observations of 
FU Ori stars, which are young stellar objects that demonstrate rapid increases in brightness over short timescales \citep{1977ApJ...217..693H}. These flares are interpreted as  boosts in the accretion rate onto the star, which can be explained by their accretion discs becoming gravitationally unstable \citep{1996ARA&A..34..207H,2016SciA....2E0875L}. 
Analogously, we expect that the accretion episodes described in this work should occur with an initially high luminosity that decreases as a portion of the gas is converted into stars.  In this case the assumption of a steady-state, constant $\dot{M}$ disc model is not necessarily valid over long timescales, for which the accretion rate will increase once the instability is triggered, then decrease as gas is depleted. Future work will explore how the time dependence of the accretion rate affects the disc geometry and stellar outcomes.

Our work supports the idea that accretion discs in AGN can boost the expected EMRI rate, not only by orbit capture of stars and BHs in the nucleus \citep{PanYang2021a,PanYang2021b}, but also by in-situ fragmentation. 
While we consider in-situ EMRIs for a range of SMBH masses, only a subset of these will be detectable by future milliHz detectors. For LISA, the sensitivity lies in the $\sim10^{-4} - 1 \rm Hz$ range (neglecting the details of detector noise as a function of frequency). A circular inspiral generates GWs at a frequency that is twice the orbital frequency, which is at most 
$f_{\rm GW} = {\Omega}/{\pi} 
\approx 4\times10^{-3} {\rm Hz} \,
({M_{\rm BH}}/{10^6 {\rm M_{\sun}}})^{1/2}
({3 r_{\rm S}}/{r})^{3/2}
$
for the final inspiral around a non-spinning SMBH. For the higher SMBH mass of $10^8 \rm M_{\sun}$, a circular EMRI will merge at lower frequencies which fall outside of the mHz band.
LISA will be ideal for detecting EMRIs around less massive host SMBHs, for which in-situ generation is critically dependent on whether stars survive disruption. In models where stars grow via a limited Bondi accretion prescription (reaching modest stellar masses $\sim10 {\rm M_{\sun}}$), this occurs at a rate that is comparable to that expected for EMRIs formed via dynamical estimates. 

Gas-embedded inspirals present an exciting class of GW sources for future milliHz detectors. Whether they are captured by the disc via orbit intersections or born in the environment in situ, their environmental evolution leads to distinguishable orbital characteristics: These events will have typically lower eccentricity compared to EMRIs formed via purely dynamical processes and possible spin alignment with the central SMBH (depending on the alignment of both constituent BHs with the accretion plane).   If several events are detected, the population properties will be skewed by their accretion and migration history.
 Individual events also present the exciting possibility of detecting signatures of their environment in the GW signal \citep{Kocsis2011,Yunes2011,Barausse2014,Derdzinski2019,Derdzinski2021,Zwick2021,LISAastro2022}, which will allow us to constrain properties of AGN discs in regions that are electromagnetically unresolvable.

\subsection{Caveats and future considerations}
There are many aspects of the picture that deserve further exploration.

Our model assumes a simplified, marginally stable disc where $Q=Q_0$, which implicitly assumes that some feedback mechanism supports the disc against further collapse. In reality, $Q$ will have radial and temporal dependence that must be solved consistently with various forms of feedback, e.g. from star formation or accretion onto embedded objects. 
(see e.g. the feedback-dominated accretion model by \citealt{Gilbaum2021}). 
This also means that the SF efficiency should have a radial dependence (e.g. as in the model of \citet{TQM2005} or the simulations by \citealt{Mapelli2012}), which will affect the resulting stellar mass distribution. Such considerations may be critical for understanding the SF in discs around less massive SMBHs, for which fragmentation occurs within the transition from optically thick to thin gas (see the opacity profile of the $10^6 \rm M_{\sun}$ disc model).

Simulations of fragmenting proto-planetary discs by \citet{Boley2010} and \citet{ChaNayakshin2011} find that clumps form on eccentric orbits. If stars are born on initially eccentric orbits, this will affect the subsequent accretion rate (more so if it is dependent on relative velocity with the local gas, as in Bondi) and orbital evolution. We plan to consider in more detail the initial conditions of stellar orbits and their effect on subsequent evolution in future work.

Furthermore, the stellar distribution in the disc that arises in our model will be relevant only for a limited time. As the gas content of the disc
diminishes due to fragmentation and accretion onto the fragments and the central AGN, stars and BHs can still evolve as a collection of
objects mutually interacting via gravity \citep{Boley2010,Zhu2012}, essentially as a collisional system. Two-body relaxation and resonant relaxation would occur, as they
do in star clusters, and potentially generate more EMRIs. This will have to be investigated in a future work, though, as the phase space structure
of the collisional disc that would have to be modelled is not a trivial extension of the case of star clusters, which is widely studied in the
literature.

We have considered only one stage of gravitational instability. 
Over longer timescales, new gas could flow into the nucleus and replenish the outer regions of the AGN disc, triggering another episode of fragmentation. 
This will be more likely
to happen at higher redshift, as more frequent interactions between the host galaxy and smaller galactic companions will trigger gas inflows \citep{2011ApJ...729...85C,2015MNRAS.447.2123C}. Dynamical instabilities in the nuclear region of the host galaxy, such as nuclear bars, could
also trigger repeated episodes of gas inflows \citep{HopkinsQuataert2011,ShlosmanFrankBegelman1989}. At a minimum, such episodes would occur on a duty cycle of order the orbital time of the
nuclear region. The latter timescale is in the range $10^6-10^7 \rm yr$ at scales of $100$ pc $- 1$ kpc for a gas-rich galaxy (e.g. \citealt{2016MNRAS.458..816H}).
Hence our EMRI rate estimate should be taken as a conservative estimate, as several cycles of in-situ star formation such as that modelled
here could occur during the lifetime of a massive gas-rich galaxy. It is only between these cycles of gas replenishment that the system will evolve
as a collisional gravitational system with a lesser role of gas.

Finally, a word of caution should also be spent about the way migration is treated in our model. Indeed, we are assuming the standard picture of linear torques, but it is known that disc-driven torques are considerably more complicated. For example, stochastic torques arise in both self-gravitating discs \citep{BaruteauMeru2011} and turbulent, magnetized discs \citep{2005A&A...443.1067N}. 
Migration rates can also be affected by the local thermal feedback from the star or an accreting BH \citep{Hankla2020,VelascoRomero2020} or mechanical feedback from stellar winds   \citep{Gruzinov2020,LiChang2020}. Furthermore, changes to the initial mass distribution, such as those resulting from the inclusion of magnetic fields (e.g. \citealt{Deng2020} will affect the subsequent migration rates.

\section{Conclusions}
We present self-consistent models of geometrically thin, steady-state, and near-Eddington accretion discs around SMBHs. By calculating the disc structure and corresponding cooling rates, we determine the conditions for fragmentation and resulting protostar properties. 
For a set of SMBH masses, disc viscosities, accretion rates, and migration estimates, we quantify the initial stellar mass distribution and a resulting EMRI rate. Our results are summarised as follows:
\begin{enumerate}
    \item For the parameters considered, discs become gravitationally unstable at sub-parsec distances. Rapid cooling and fragmentation into protostars occurs at a few $\times 10^{-3} - 10^{-2} \rm pc$ from the central SMBH, and extends up to $r\sim 0.1 \rm pc$. 
    \item Initial stellar mass distributions are top-heavy and peak at $0.1-1 {\rm M_{\sun}}$ for discs around $M_{\rm BH} = 10^6 {\rm M_{\sun}}$, $1-10 {\rm M_{\sun}}$ for $M_{\rm BH} = 10^7 {\rm M_{\sun}}$, and $10-10^2 {\rm M_{\sun}}$ for $M_{\rm BH} = 10^8 {\rm M_{\sun}}$. Assuming a SF efficiency of $1$ percent, fragmentation results in $N_*\gtrsim 10^2 - 10^3$ stars with a relatively flat mass distribution slope of $dN/dm \propto M_*^{-0.7}$. 
    \item Subsequent accretion onto stars via a limited Bondi prescription (which preserves the steady state assumption of the global accretion disc) results in stars increasing their masses by $\sim2$ orders of magnitude over the AGN disc lifetime of $\sim10^7 \rm yr$. 
    \item We calculate the timescale for stars to migrate from their initial positions to the $10$ Schwarzschild radii, where they are either tidally disrupted or accreted by the central SMBH.  Migration of the stellar population leads to three outcomes: tidal disruption of stars, in-situ extreme mass ratio inspirals with the central SMBH, or leftover stars at sub-parsec separations from the central SMBH. 
    \item The rate of in-situ EMRIs produced via the orbital evolution of these stellar populations is $R_{\rm emri}\sim 0- 10^{-4} \rm yr^{-1}$ per AGN per SMBH accretion episode. Not all models considered produce EMRIs: for $M_{\rm BH}=10^6 {\rm M_{\sun}}$, stars need to survive tidal disruption, so EMRIs only occur when stars are able to grow to $\gtrsim 10 {\rm M_{\sun}}$. Our rate is an under-estimate for SMBHs that experience multiple activity cycles.
    \item EMRI rates are dependent on the SF efficiency, stellar accretion, and migration prescription. Rates will increase if star formation produces a higher number of massive stars, or if deviations from migration occur that slow down the migration of stars towards radii where they become victim to tidal forces. 
    \item Our disc models suggest the presence of migration traps due to changes in density and temperature gradients. These occur at $\sim50-300 r_{\rm S}$, depending on the strength of viscosity ($\alpha$). At smaller separations, the orbital evolution of stars/BHs in these regions will also be governed by GW emission, and thus the traps only slow down their inevitable inspiral. For the more distant traps, the delay in inward migration may be more substantial. 

\end{enumerate}

AGN-born star populations may be diverse depending on the host SMBH and the gas supply, suggesting that this process will vary throughout cosmological history as SMBHs grow.

\section{Acknowledgements} 
We are grateful to the referee, Doug Lin, for his insightful comments that improved this work. 
We thank Zolt\'an Haiman, Nicholas Stone, and Alex Dittmann for comments on the manuscript and 
Tassos Fragos, Lorenz Zwick, Mudit Garg, and Simona Pacuraru for inspiring conversations. 
The authors acknowledge support from the Swiss National Science Foundation 
Grant 200020\_192092. AD acknowledges support from the Tomalla Foundation for Gravity Research. We also acknowledge use of the following \emph{software}: NumPy \citep{numpy}, SciPy \citep{2020SciPy-NMeth}, and Matplotlib \citep{Hunter2007}.

\section{Data availability statement}
The calculations and data used in this study are available upon request from AD.

\bibliographystyle{mnras}

\bibliography{frag.bib}

\appendix

\section{Migration traps}
\label{sec:traps}
With our disc solutions we can calculate the torque on embedded objects with an analytical fit for the Type I torque that takes into account the local density, temperature, and entropy gradients of the disc. These estimates are tested in simulations for the case of fully unsaturated co-rotation torque (relevant for discs with high viscosity, \citealt{PP2009}), in the non-isothermal \citep{KleyCrida2008} and isothermal \citep{Tanaka2002,Paardekooper2010} limits. 
In this case the Type I torque in the locally isothermal regime is given by \citep{Paardekooper2010}
\be
T_{\rm iso}/T_0 = 0.85 - \alpha_T - 0.9 \beta_T 
\ee
where $\alpha_T = - \frac{d \ln{\Sigma}}{d \ln{r}}$, $\beta_T = - \frac{d \ln{T_{\rm mid}}}{d \ln{r}}$,
and the normalization is  
\be
T_0 = q^2 \left(\frac{r}{h} \right)^2 \Sigma r^4 \Omega^2,
\ee
where $\Omega$ is the orbital frequency of the migrator, $q=M_*/M_{\rm BH}$ is the mass ratio, and the disc properties correspond to those at the migrator's position $r$.
In the adiabatic limit, the torque follows 
\be
\gamma T_{\rm ad}/T_0 = 0.85 - \alpha_T - 1.7 \beta_T - 7.9 \xi / \gamma 
\ee
where $\xi = \beta_T - (\gamma - 1)\alpha_T$.  
Following \citet{Lyra2010}, we interpolate between the two torque regimes to obtain the total torque:
\be
T_{\rm total}/T_0 = \frac{T_{\rm ad} \beta_{\rm crit}^2 + T_{\rm iso}}{(1 + \beta_{\rm crit})^2}
\ee
where $\beta_{\rm crit} = t_{\rm cool}/t_{\rm dyn}$ is the ratio of the radiative cooling timescale to the dynamical timescale, $t_{\rm dyn} = 2 \pi / \Omega$, and $t_{\rm cool}$ is defined in Sec~\ref{sec:discmodel}.

\begin{figure}
\begin{center}
\includegraphics[width=0.45\textwidth]{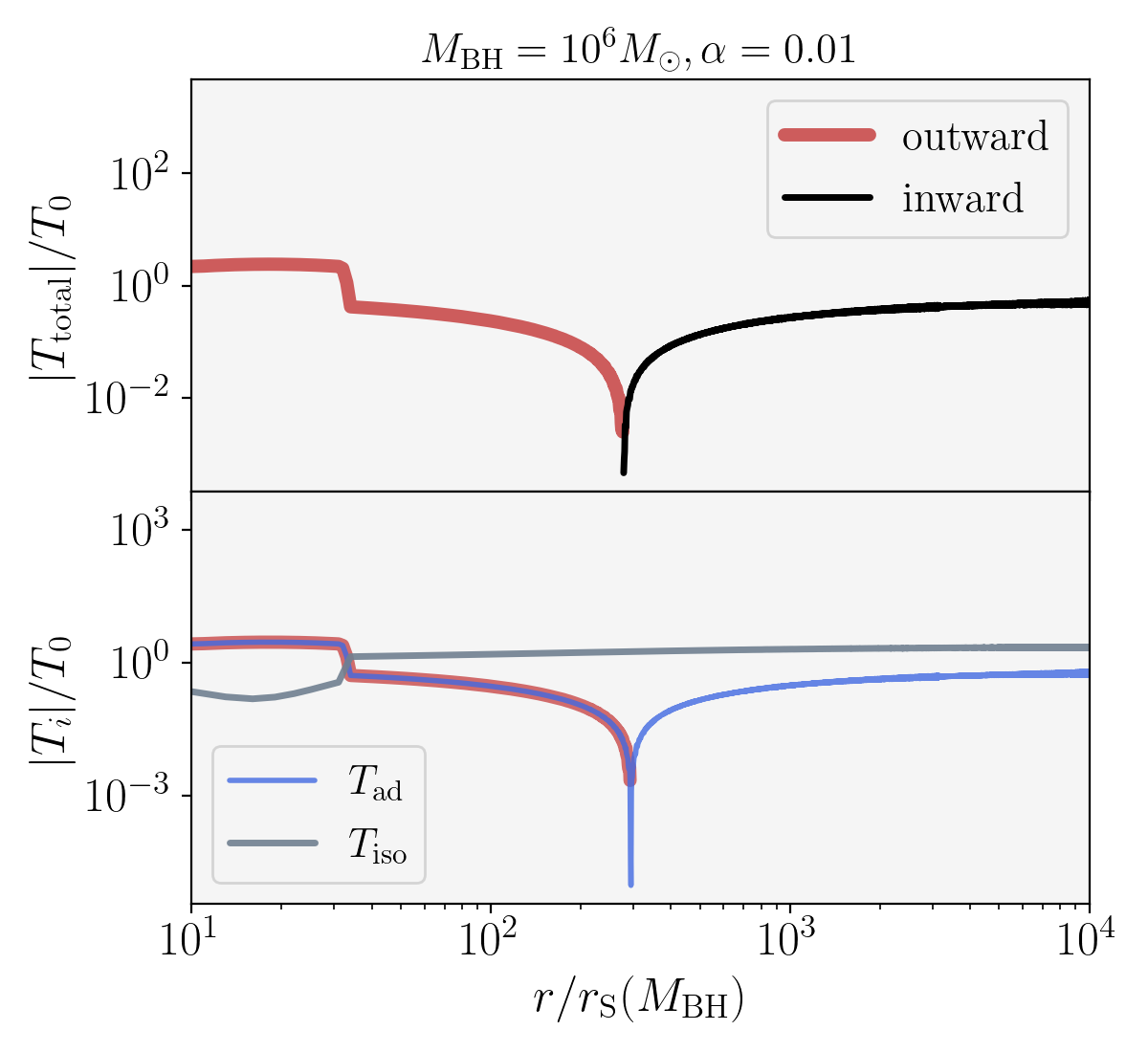}\\
\vspace{-2mm}
\includegraphics[width=0.45\textwidth]{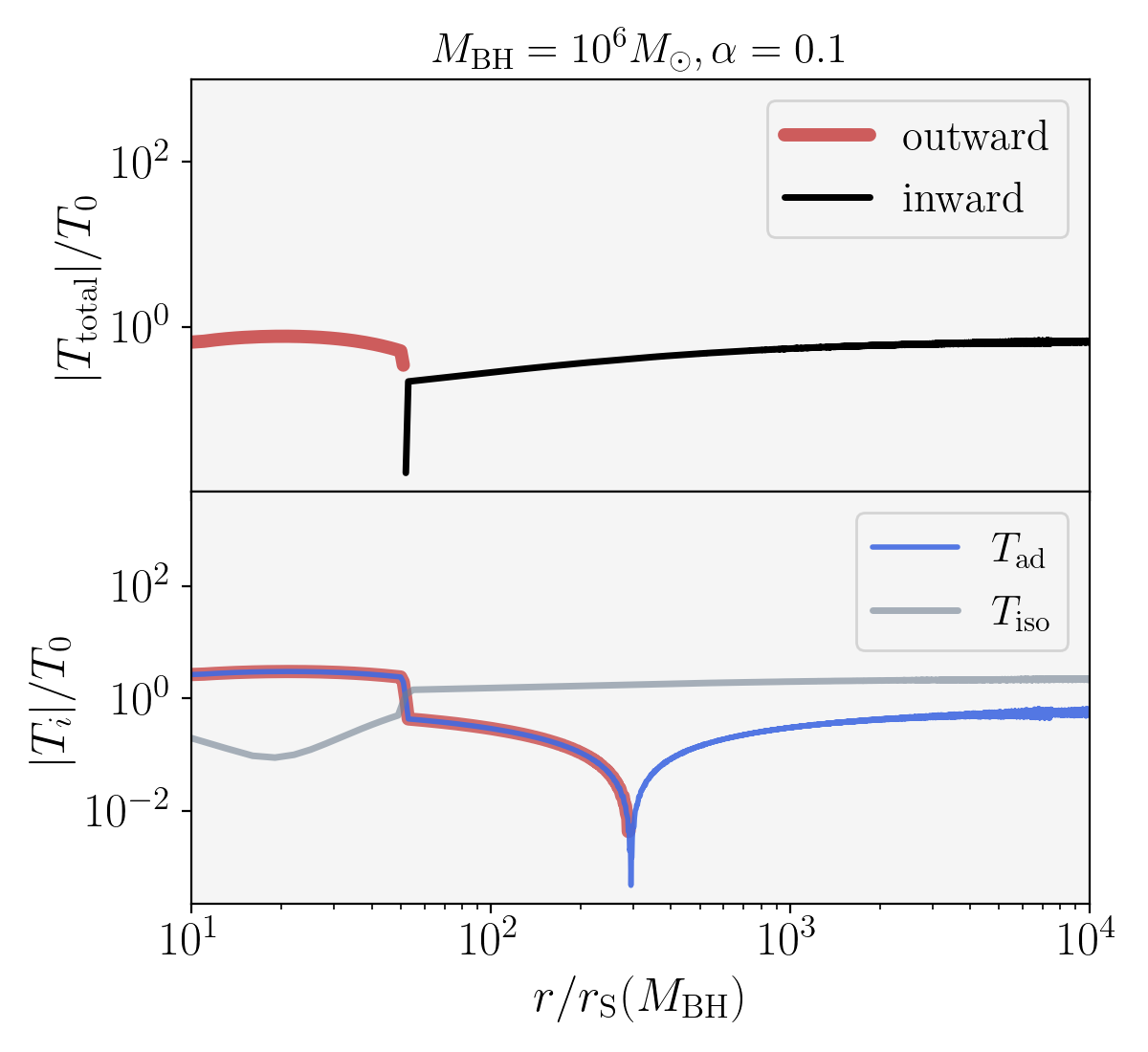}\\
\vspace{-2mm}
\caption{ The absolute value of the Type I torque profiles normalized by $T_0$. Red highlighted regions show where the torque is positive (outward) due to changes in the density and temperature gradients. The total torque (solid black line) is a combination of adiabatic and isothermal, depending on the disc cooling rate. [{\it Top panel}:] For M6 and $\alpha=0.01$: In the inner regions where gas is optically thick, radiative timescales are long and migration is in the adiabatic regime, resulting in a migration trap at $\sim 300 r_{S}$.
[{\it Bottom panel}]: M6 and $\alpha=0.1$. the disc density is lower, so the total torque is in between the adiabatic and isothermal regimes. The migration trap occurs at smaller radii ($\sim 50 r_{\rm S}$) due to changes in the cooling rate. The location of migration traps depends on $\alpha$.
}
\label{fig:torques}
\end{center}
\end{figure}

In Fig.~\ref{fig:torques} we plot this Type I torque as a function of radius for the isothermal, adiabatic, and interpolated regimes. While the torque is typically negative (inward), there are regions where the torque becomes positive (outward) due to changes in the disc density and temperature gradients. 
In regions where the cooling is rapid ($\beta_{\rm crit}<<1$), the torque converges to the isothermal assumption. Conversely in regions where the the optical depth is high, the torque is in the adiabatic regime. $T_{\rm ad}$ is typically weaker than the $T_{\rm iso}$ at intermediate radii, but becomes stronger and positive at inner radii in regions where radiation pressure becomes important.

The transition from negative (inward) to positive (outward) torque suggests the presence of migration traps in the inner disc, or regions where torques cancel and several migrators may congregate, as discussed in \citep{Bellovary2016, Lyra2010}.
We caution that studies adopted to identify these thermodynamical effects on the torques are based on simulations of non-self-gravitating discs. 
 However, outward migration has been observed also in self-gravitating discs, especially in adiabatic conditions \citep{2007prpl.conf..607D,SzulagyiMayer2017}.
We note that similar disc models by \citet{PanYang2021a} claim that no migration traps occur--however, the models considered in these works do not include the adiabatic component of the torque, which leads to a torque transition in the inner disc region. Their work also demonstrates that the inclusion of other torque components (such as from an accretion head wind, not considered here), may also affect the robustness of migration traps.

Traps do not always halt the migration of embedded stars, if gravitational wave emission becomes relevant in the regions where traps occur. 
While the gas torque becomes weaker at smaller separations, the angular momentum loss due to GW emission becomes stronger, eventually comparable to or greater than the gas torque ($\dot{r}_{\rm GW} > \dot{r}_{\rm gas}$ at a few $\times 10 r_{\rm S}$, depending on the binary mass). Thus the inclusion of GW emission over sufficiently long timescales can eventually negate any long-term effect of migration traps. In cases where traps occur around $50 r_{\rm S}$,  rather than being `traps,' they are regions where the gas torque changes sign and migration \emph{stalls}. The overall inward migration slows down but continues and is eventually dominated by GWs. 

Even if traps only slow down inward migration, their effect has more significance in cases where stars are vulnerable to subsequent disruption. Traps occur outside the regime where tidal forces will destroy embedded stars. For the $M_{\rm BH} = 10^6 \rm M_{\sun}$ disc model, this has implications for the number of stars that seed potential EMRIs. In this case traps may aide the survival of embedded stars, especially in cases where they occur at larger radii ($\sim300 r_{\rm S}$ for $\alpha=0.01$), where the GW emission is weak ($\dot{r}_{\rm GW}(300 r_{\rm S})/\dot{r}_{I}\sim10^{-5}$). This would lead to a delay in migration that could allow stars to survive and produce CRs, increasing the EMRI rate. We leave the quantification of this to future work, since one should also consider the mutual interaction of stars that will occur if they congregate in migration traps.

\begin{figure}
\begin{center}
\includegraphics[width=0.45\textwidth]{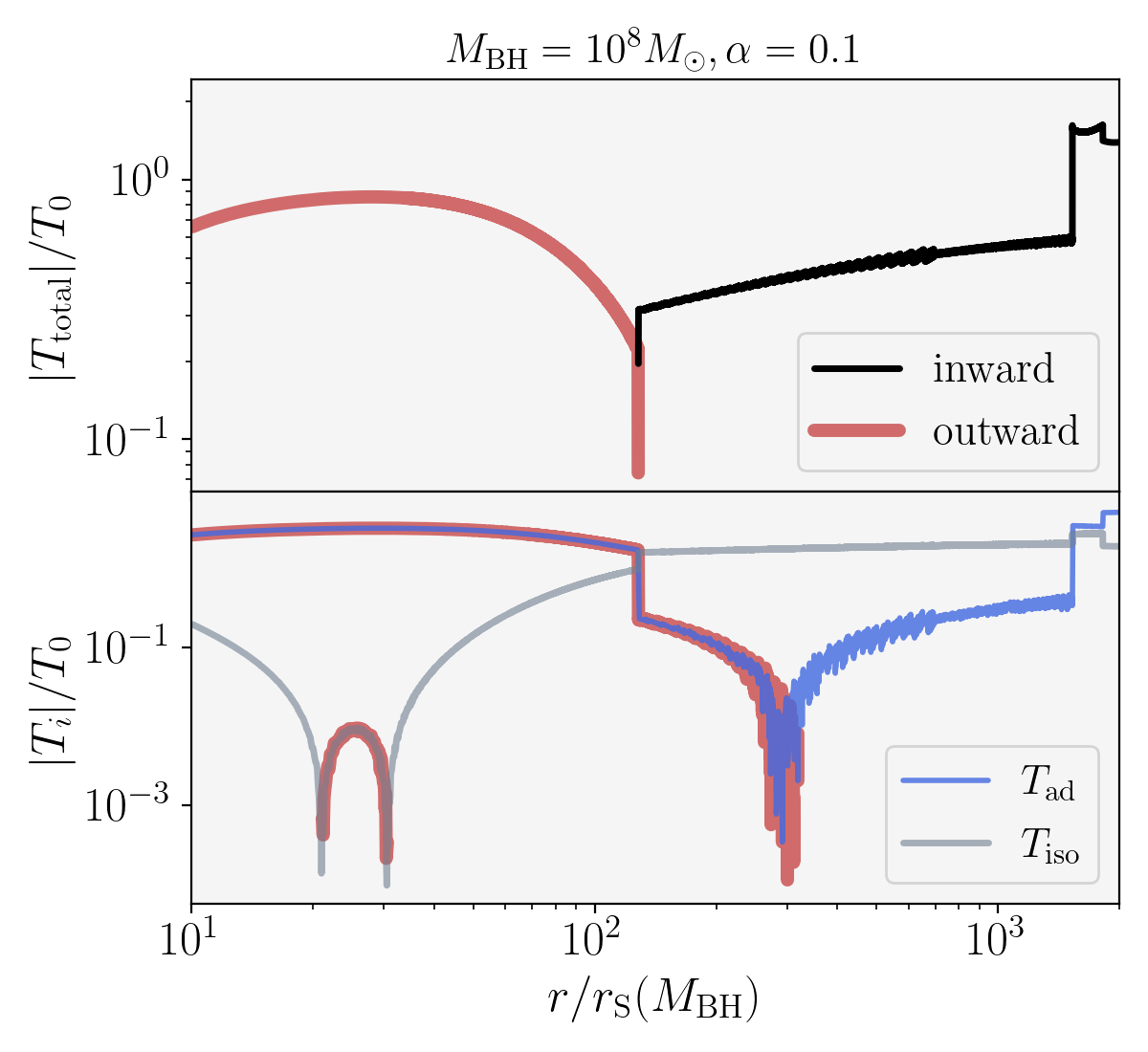}\\
\vspace{-2mm}
\includegraphics[width=0.45\textwidth]{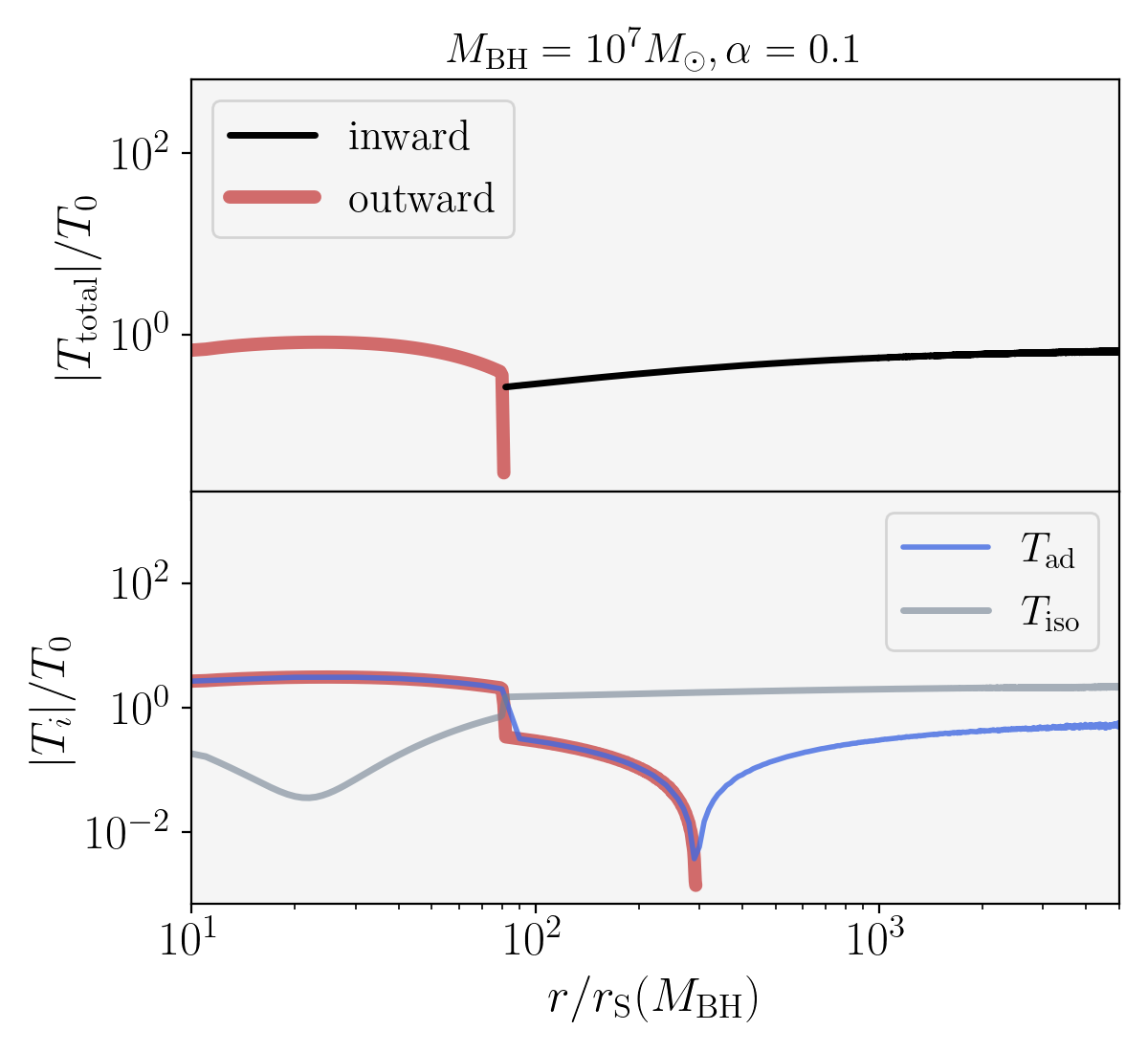}\\
\vspace{-2mm}
\caption{ Absolute value of the total, isothermal, and adiabatic torque profiles for $M_{\rm BH} = 10^8 {\rm M_{\sun}}$ and $M_{\rm BH} = 10^7 {\rm M_{\sun}}$ for discs with $\alpha=0.1$. Migration traps occur at $\sim 130 r_{\rm S}$ (for $10^8 \rm M_{\sun}$) and at $\sim80 r_{\rm S}$ (for $10^7 \rm M_{\sun}$).}
\label{fig:moretorques}
\end{center}
\end{figure}

\bsp	
\label{lastpage}
\end{document}